\documentclass[useAMS,usenatbib]{mn2e}

\usepackage{graphicx}
\newcommand{\pbj}{\ifmmode{P_{\rm B1}}\else$P_{\rm B1}$\fi}
\newcommand{\pbd}{\ifmmode{P_{\rm B2}}\else$P_{\rm B2}$\fi}
\newcommand{\pbt}{\ifmmode{P_{\rm B3}}\else$P_{\rm B3}$\fi}
\newcommand{\fbj}{\ifmmode{f_{\rm B1}}\else$f_{\rm B1}$\fi}
\newcommand{\fbd}{\ifmmode{f_{\rm B2}}\else$f_{\rm B2}$\fi}
\newcommand{\fbt}{\ifmmode{f_{\rm B3}}\else$f_{\rm B3}$\fi}

\title[Blazhko-type modulation in double-mode RR~Lyrae stars]{Blazhko-type modulation in the double-mode RR~Lyrae stars of the OGLE Galactic bulge collection}
\author[R. Smolec et al.]
{R. Smolec$^{1}$\thanks{E-mail: smolec@camk.edu.pl}, I. Soszy\'nski$^{2}$, A. Udalski$^{2}$, M.K. Szyma\'nski$^{2}$, P. Pietrukowicz$^{2}$\and
J. Skowron$^{2}$, S. Koz\l{}owski$^{2}$, R. Poleski$^{2,3}$, D. Skowron$^{2}$, G. Pietrzy\'nski$^{2,4}$\and
\L{}. Wyrzykowski$^{2,5}$, K. Ulaczyk$^{2}$ \& P. Mr\'oz$^{2}$ \\
$^{1}$ Nicolaus Copernicus Astronomical Centre, Polish Academy of Sciences, Bartycka 18, 00-716 Warszawa, Poland\\
$^{2}$ Warsaw University Observatory, Al. Ujazdowskie 4, 00-478 Warszawa, Poland\\
$^{3}$ Department of Astronomy, Ohio State University, 140 W. 18th Ave., Columbus, OH 43210, USA\\
$^{4}$ Universidad de Concepci\'on, Departamento de Astronomia, Casilla 160-C, Concepci\'on, Chile\\
$^{5}$ Institute of Astronomy, University of Cambridge, Madingley Road, Cambridge CB3 0HA, UK
}

\begin{document}

\date{Accepted . Received ; in original form }

\pagerange{1--21} \pubyear{2014}

\maketitle

\label{firstpage}

\begin{abstract}
We present the analysis of Blazhko-type modulation in double-mode RR~Lyrae (RRd) stars found in the OGLE photometry of the Galactic bulge. Modulation is detected in 15 stars. Most of them have non-typical period ratio of the radial modes. In the Petersen diagram, at a given period of the fundamental mode, they are located significantly below or above the sequence formed by the majority of RRd stars. 

Multi-periodic modulation is very frequent; two or three modulation periods are detected in 8 stars. Modulation periods vary from $\sim$20 to more than 300\thinspace days. Radial mode amplitudes can be modulated by a few to nearly hundred per cent. Both radial modes may be modulated with the same period. More commonly however, dominant modulation for the fundamental mode has different period than dominant modulation for the first overtone. Quite often modulation of only one mode is detected in the data. We find a clear feedback between pulsation amplitude of the dominant mode and mean stellar brightness: lower the pulsation amplitude, brighter the star. At phases of low pulsation amplitude, the mode periods are prone to fast changes. All the stars share the common feature: their pulsation properties are non-stationary. Amplitudes and phases of the radial modes vary irregularly on a long time-scale of a few hundred or thousand days. The short-term modulations are also irregular.

One of the stars has switched the pulsation mode recently: from single-mode fundamental mode pulsation to RRd state. In other star the non-radial mode with characteristic $\sim\!0.61$ period ratio to the first overtone is detected. This non-radial mode is likely modulated with the same period as the radial modes.
\end{abstract}

\begin{keywords}
stars: horizontal branch -- stars: oscillations -- stars: variables: RR~Lyrae
\end{keywords}

\section{Introduction}\label{sec:intro}

RR~Lyrae stars are classical pulsators oscillating in the radial modes. A majority of these stars are single-mode pulsators pulsating either in the fundamental mode (F-mode, RRab stars) or in the first overtone mode (1O-mode, RRc stars). Less frequent is pulsation in the two modes simultaneously (RRd stars). RR~Lyrae stars play an important role in distance determination and in the studies of Galactic structure  and evolution. RRd stars, or multiperiodic pulsators in general, are important for the determination of basic stellar parameters, in particular of stellar masses \citep[see e.g.][]{petersen,popielski}.

RR~Lyrae stars are relatively simple pulsators, nevertheless they are not well understood. The most stubborn problem is the origin of the Blazhko effect -- a long term quasi-periodic modulation of pulsation amplitude and phase \citep{blazhko}. The other unsolved problem is mode selection mechanism. We do not know why some stars pulsate in two modes simultaneously and what is the mechanism behind \citep[for a review see][]{smolec14}. In the recent years our knowledge about RR~Lyrae stars significantly increased. The progress is not only thanks to space photometry revolution, caused by space telescopes {\it CoRoT} and {\it Kepler}, but also thanks to long-term ground-based photometric sky surveys, like the Optical Gravitational Lensing Experiment (OGLE).

 The ultra-precise photometry of space telescopes allows to investigate a relatively small sample of stars, but with unprecedented accuracy and detail. As a result our knowledge about the Blazhko effect significantly increased. Space photometry indicates that nearly $50$ per cent of RRab stars are affected by the Blazhko effect \citep{benko10,benko14}. The most important discoveries are: detection of period doubling effect in about half of the modulated stars observed from space \citep{szabo10,szabo14}, detection of additional radial modes in the Blazhko variables in the mmag regime \citep[mainly of the second overtone,][]{benko10,benko14,molnar12}, multiple modulation periods and clear demonstration of irregular nature of the modulation \citep[e.g.][]{benko14,gugg}.

Although new models behind the Blazhko modulation were proposed \citep[e.g.][]{bk11} we are still far from the solution of the Blazhko puzzle \citep[for a review see][]{szabo14}.

Space photometry allowed clearly establishing a new group of double-mode pulsators, pulsating in the first overtone and in a yet not identified non-radial mode, with characteristic period ratio of the two modes in a narrow range around $P_x/P_1\approx 0.61$. Such period ratios were also detected in RRd stars. First such stars were discovered in the {\it MOST} photometry and in the ground-based photometry of $\omega$~Centauri \citep{aqleo,om09}. Analysis of space observations of RRc and RRd stars indicates that this form of pulsation must be common (e.g. \cite{chadid12}, \cite{pamsm14} and references therein, \cite{szabo_corot}). Nearly all RRc/RRd stars observed from space show the additional non-radial mode. A similar form of pulsation was detected in the first overtone Cepheids, mostly in the OGLE data \citep{mk09,ogle_cep_smc}. We do not understand this form of pulsation \citep[see however][]{wd12}.

Many interesting discoveries come from analysis of data collected by the long-term ground based photometric campaigns, of which the Optical Gravitational Lensing Experiment (OGLE) is the longest and of highest photometric quality \citep[e.g.][]{ogleIII}. The large area of sky monitored by the OGLE project allowed identification of several hundred thousands of pulsating stars only during the first three phases of the experiment. More than $38\,000$ of RR~Lyrae stars were identified in the OGLE-III and OGLE-IV photometry of the Galactic bulge \citep{ogleiv_rrl_blg}. Such a large collection of RR~Lyrae stars must contain unique objects. Indeed, a few mode switching stars were reported \citep{ogle_switch,ogleiv_rrl_blg}, a gravitationally lensed RRc star was observed \citep{ogleiv_rrl_blg}, and new type of pulsator mimicking RR~Lyrae variability, binary evolution pulsator, was discovered \citep[BEP;][]{gp12,smolec_bep}. The OGLE data is also perfect for search and study of non-radial modes of low amplitude. The study of Galactic bulge OGLE-III data by \cite{netzel} increased a sample of known radial--non-radial double-mode pulsators by a factor of $6$. Yet another application is study of the Blazhko effect \citep[e.g.][]{mizerski,mp03}.

In this paper we present the analysis of the Blazhko-type modulation detected in several RRd stars from the OGLE-III and OGLE-IV photometry of the Galactic bulge \citep{ogle_rrl_blg,ogleiv_rrl_blg}. These are the first objects of this type, in which modulation is observed on top of the genuine double-mode pulsation. The properties of the stars are studied in detail. We note that after \cite{ogleiv_rrl_blg} announced the discovery of the first modulated RRd stars, \cite{jurcsik_bl} also reported the modulated RRd stars in their photometry of M3 (see also Sec.~\ref{ssec:dis_other}).

Structure of the paper is the following. In Section~\ref{sec:methods} we present the data and the adopted analysis methods. The general overview of the data with basic properties of double-mode pulsation and properties of modulation are presented in Section~\ref{sec:overview}. In Section~\ref{sec:individual} we discuss the analysis of individual stars in detail. Discussion of the emerging picture of the modulation in RRd stars is in Section~\ref{sec:discussion}. Summary closes the paper.

\section{Observations and data analysis}\label{sec:methods}
The I-band light curves we analyse in this paper were collected during the third and fourth phases of the Optical Gravitational Lensing Experiment. The OGLE-III project was conducted from 2001 to 2009. The number of data points in the OGLE-III light curves varies strongly from field to field: from 335 to 2933. OGLE-IV is an ongoing survey started in March 2010. In this study we have used the data obtained up to September 2014, so the time span of the OGLE-IV observations reaches 4.5 years, with the total number of the observational points from about 1000 to 10000. The accuracy of the individual photometric measurements is better than 10\thinspace mmag for most of the stars in our sample. The reader is referred to  \cite{ogleIII} and \cite{ogleiv_rrl_blg} for description of the instrument setup and photometry reduction procedures. The photometry is publicly available and may be downloaded from the OGLE ftp archive \citep[\textsf{ftp://ftp.astrouw.edu.pl/ogle/ogle4/OCVS/blg/rrlyr/}; for the description see][]{ogleiv_rrl_blg}.

The data were analysed using two techniques. The first is a standard successive prewhitening of the data with frequencies detected in the discrete Fourier transform of the data. At each step the data are fitted with the sine series of the following form:
\begin{equation}
m(t)=m_0+\sum_{k=1}^N A_k\sin\big(2\pi f_k t + \phi_k\big)\,,\label{eq:ss}
\end{equation}
where $f_k$ are independent frequencies detected in the data and their possible linear combinations. Amplitudes, phases and frequencies are adjusted using the non-linear least square fitting procedure.  We regard two frequencies, $f_{\rm a}$ and $f_{\rm b}$, as unresolved if distance between them $|f_{\rm a}-f_{\rm b}|<2/T$, where $T$ is data length. Only resolved signals are included in eq.~(\ref{eq:ss}).

For all analysed stars we first find {\it full light curve solution}: in the frequency spectrum of the residuals we do not detect any significant signal, or, what is more common, we can detect only unresolved power at the prewhitened frequencies, and possibly a signal at low frequencies -- a signature of slow trends. At this phase the data are $6\sigma$ clipped, the trend is removed from the original data with polynomial of chosen order and again, light curve solution is updated and outliers at a lower $5\sigma$ level are removed. All further analysis is conducted on the resulting detrended time series with severe outliers removed.

In a majority of the cases, frequencies included in the full light curve solution are: frequencies of the fundamental mode and of the first overtone, $f_0$ and $f_1$, respectively, their linear combinations, $kf_0\pm lf_1$, and modulation components arising form the detected Blazhko effect. These frequencies appear as equally spaced multiplet structures at $f_0$, $f_1$ and their combinations, or as close doublets (incomplete triplets, see next Section for more details). The separation between the multiplet (doublet) components defines the modulation (Blazhko) frequency, $f_{\rm B}$, and modulation period, $P_{\rm B}=1/f_{\rm B}$. If the modulation period is longer than $T/2$, or pulsation mode (its amplitude and/or phase) vary on a long time-scale, unresolved power will remain in the frequency spectrum at $f_0$ and/or at $f_1$ after prewhitening. In the case of OGLE-IV data $T\approx\! 1600$\thinspace d and the longest possible modulation we can detect is  $\approx\! 800$\thinspace days. In the combined OGLE-III and OGLE-IV data this limit is longer, up to $\approx\! 3000$\thinspace days (depending on the time span of OGLE-III observations). 

By default the analysis is based on OGLE-IV data which are more densely sampled. When we suspect long-term modulation/variation of the radial modes, and OGLE-III data are available, we also analyse the merged data. We note that there are systematic differences in the zero-points in OGLE-III and OGLE-IV data, which may be as high as $0.2$\thinspace mag in extreme cases \citep{ogleiv_rrl_blg}. To correctly merge the data we first find the full light curve solution for OGLE-III and for OGLE-IV data independently, as described above. The mean brightness difference, $\Delta m_0$, between OGLE-III and OGLE-IV data is then taken into account when merging the data. 

The side peaks at the radial mode frequencies can also be non-coherent, i.e. they cannot be prewhitened with a single sine wave with constant amplitude and phase. After prewhitening, residual, unresolved power remains in the spectrum. The presence of such signals is not a surprise -- they may originate from the non-periodic nature of the light variation, which is expected. The Blazhko effect in single periodic RR~Lyrae stars is a quasi-periodic effect. The results of space missions leave no doubt: consecutive Blazhko cycles differ, modulation periods and amplitudes sometimes change in an irregular fashion \citep[e.g.][]{benko14,gugg}. 

To analyse the modulation and/or possible long-term variation of the radial modes in more detail, we adopted the second technique, the time-dependent Fourier analysis \citep{kbd87} in some cases followed by time-dependent prewhitening \citep{pamsm14}. In a nutshell, the data are divided into chunks or subsets of approximately the same length, $\Delta t$. Then, for each of the subsets we fit the sine series (eq.~\ref{eq:ss}) with $f_0$, $f_1$, and their linear combinations detected in the full data set. These frequencies are kept fixed, i.e. are the same for all subsets -- we adjust amplitudes, phases, and mean brightness only. Any variation of mode frequencies is then reflected in the variation of respective phases. The difference between the instantaneous and the mean frequency is:
\begin{equation}
\Delta f=\frac{1}{2\pi}\frac{d\phi}{dt}\,.\label{eq:pc}
\end{equation}
Using this method we can study the variation of mode's amplitudes and periods on time-scale longer than $\Delta t$. For a few stars with OGLE-IV photometry the data sampling is dense enough to use $\Delta t$ as low as $5$\thinspace days. To characterise the data sampling we calculate the mean number of data points per $100$\thinspace days, $\rho$ in the following. The gaps between observing seasons are excluded when estimating $\rho$. This parameter vary from $\approx 100$ to more than $800$ in the OGLE-IV data (typically above $300$) and from $\approx 30$ to $\approx 140$ in the OGLE-III data. 

If variation occurs on a long time-scale (of order of data length) and is irregular it produces unresolved power in the frequency spectrum at $f_0$ and/or at $f_1$, which increases the noise level in the Fourier transform and may hide e.g. the signs of modulation on the shorter time-scale. To get rid of these unwanted signal we follow the time-dependent Fourier analysis by the time-dependent prewhitening \citep{pamsm14}. The residuals from the just described sine series fits, separate for each of the subsets, are merged. In a resulting time series the long-term variability is filtered out, while short-term modulation remains. Of course $f_0$, $f_1$, and their combinations are also prewhitened and should not appear in the frequency spectrum, provided they do not vary significantly on a time-scale shorter than $\Delta t$.

All analysis described in this paper was conducted using the software written by one of us (RS).

\section{Overview of the data}\label{sec:overview}

\begin{figure*}
\centering
\resizebox{\hsize}{!}{\includegraphics{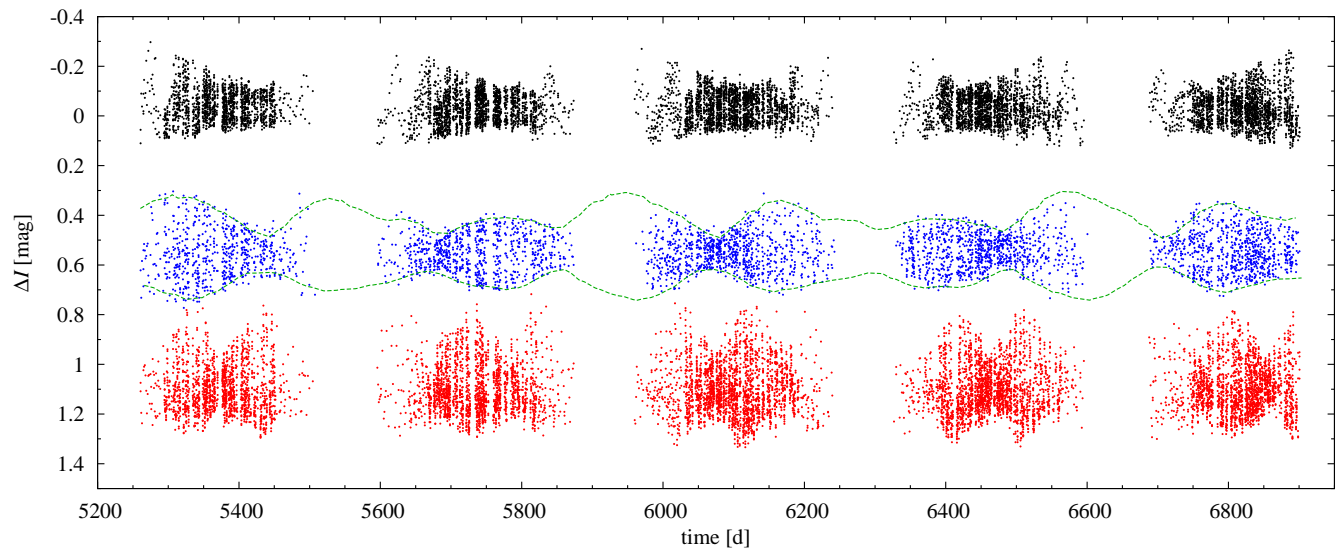}}
\caption{OGLE-IV data for the three modulated RRd stars; from top to bottom: OGLE-BLG-RRLYR-06283, -07393 and -05762 (arbitrarily shifted in the vertical direction). The dashed line plotted for OGLE-BLG-RRLYR-07393 corresponds to the envelope of the full light curve solution (see Sec.~\ref{sec:07393}). Here and in this paper time is expressed as $t={\rm HJD}-2\,450\,000$.}
\label{fig.ts}
\end{figure*}

\begin{figure}
\centering
\resizebox{\hsize}{!}{\includegraphics{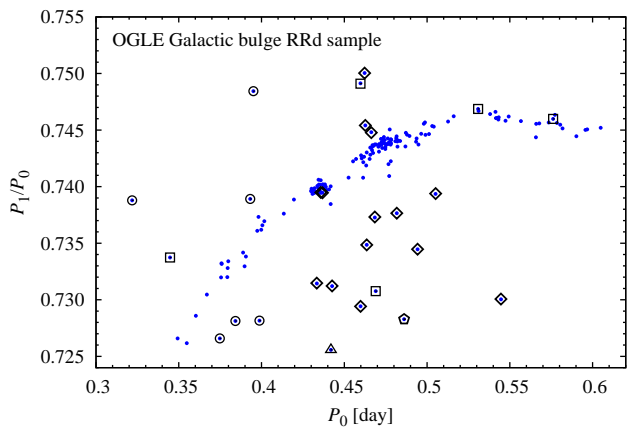}}
\caption{The Petersen diagram for 173 RRd stars of the OGLE Galactic bulge sample \citep{ogleiv_rrl_blg}. Stars that show the Blazhko-type modulation are marked with diamonds/pentagon. Location of two stars overlap at a clump corresponding to suspected members of the tidal stream crossing the bulge. In stars marked with circles we do not find close secondary peaks. In stars marked with squares a long-term variation of first overtone is present. Triangle marks the triple-mode RR~Lyrae star \citep{smolec_tri} and pentagon marks the star that recently switched the pulsation mode (and is also modulated).}
\label{fig.pet}
\end{figure}

In this study we analyse RRd stars suspected of long-term modulation of radial-modes, the Blazhko effect. The sample was first preselected by ({\it i}) visual inspection -- see Fig.~\ref{fig.ts} for obvious signature of modulation and ({\it ii}) by automatic analysis of frequency spectra of all RRd stars. In the latter case, the stars showing additional close side peaks at the fundamental and/or at the first overtone frequency were marked as suspected of long-term modulation. For details on the automatic analysis of frequency spectrum, see \cite{ogleiv_rrl_blg}. The selected sample contained 16 stars (primary targets). Our analysis showed that nearly all of these RRd stars that show the modulation, have non-typical period ratios (see Fig.~\ref{fig.pet}). Therefore, the sample was extended with 11 other stars (secondary targets) having non-typical period ratios, of which in two we detected modulation. Altogether $27$ stars were analysed in detail (marked with large symbols in Fig.~\ref{fig.pet}) and in 15 (marked with diamonds/pentagon) we found modulation of at least one radial mode.

All of the investigated stars pulsate simultaneously in the fundamental and in the first overtone modes (RRd). In all cases we detect combination frequencies of $f_0$ and $f_1$ -- no doubt the two frequencies originate from the same star. In all but two cases the period ratios $P_1/P_0$ are rather far from that typical for a majority of the Galactic bulge RRd stars -- see the Petersen diagram in Fig.~\ref{fig.pet}. The two stars that fit the main progression are also not-typical -- they are located at nearly the same place, at a clump of 28 stars -- the likely members of a tidal stream crossing the Galactic bulge, as analysed by \cite{ogleiv_rrl_blg}. We will comment the non-typical location of modulated stars in the Petersen diagram in the Discussion (Sec.~\ref{ssec:dis_pet}). In Tab.~\ref{tab:dm} we collect periods, period ratio and amplitudes of the fundamental and first overtone modes. These values come from the full light curve solution of OGLE-IV data, only, and are mean values (see Tables in the Appendix in the on-line version of the journal). The solutions include frequencies of both modes, their combinations and all additional close side peaks we could detect in the data. If the peak at $f_0$ or at $f_1$ is non-coherent, its amplitude is marked with asterisk in Tab.~\ref{tab:dm}. The period of the dominant, higher amplitude mode is written with bold fonts. In 8 stars pulsation in the first overtone dominates ($\approx 53$\thinspace per cent). In the full OGLE RRd sample (173 stars\footnote{\cite{ogleiv_rrl_blg} lists 174 RRd stars. During this study we have found that one of these stars is not RRd pulsator; the daily alias of non-coherent signal at the suspected fundamental mode frequency was misinterpreted as the first overtone.}) the dominant pulsation in the first overtone is significantly more frequent and occurs in $82.1$\thinspace per cent of the sample. We comment on this more in the Discussion.

\begin{table*}
\centering
\caption{Basic properties of RRd stars derived from the analysis of the OGLE-IV photometry. Period of the dominant mode is written with bold font. Asterisk at mode amplitude indicates that unresolved signal remains at $f_0$ or at $f_1$ in the frequency spectrum of the data prewhitened with the full light curve solution.}
\label{tab:dm}
\begin{tabular}{lrrrrrr}
star & $P_0$\thinspace [d]& $P_1$\thinspace [d] & $P_1/P_0$ & $A_0$\thinspace[mmag] & $A_1$\thinspace[mmag] & remarks\\ 
\hline
OGLE-BLG-RRLYR-00951 & {\bf 0.5446346(6)} & 0.3976183(3) & 0.7301 & *73.4 & *58.0 & \\
OGLE-BLG-RRLYR-02530 & {\bf 0.463563(1)}  & 0.3406466(7) & 0.7348 & 95.6 & *67.3 & \\
OGLE-BLG-RRLYR-02862 & 0.459812(1) & {\bf 0.3354003(6)} & 0.7294 & *50.7 & 81.4 & \\
OGLE-BLG-RRLYR-04598 & 0.4622854(6) & {\bf 0.3467316(4)} & 0.7500 & 64.9 & *71.0&$P_1\!:\!P_0=3\!:\!4$ \\
OGLE-BLG-RRLYR-05762 & {\bf 0.4662964(1)} & 0.3472844(2) & 0.7448 &  94.5 & *51.5 & \\ 
OGLE-BLG-RRLYR-06283 & 0.5051447(4) & {\bf 0.3734973(2)} & 0.7394 &  52.4 & *62.5 & \\ 
OGLE-BLG-RRLYR-07393 & 0.4626906(6) & {\bf 0.3448963(3)} & 0.7454 &  44.7 & *56.9 & $P_x/P_1=0.6163$\\   
OGLE-BLG-RRLYR-09284 & 0.4367995(4) & {\bf 0.3229961(1)} & 0.7395 & *25.5 &  60.1 & tidal stream \\ 
OGLE-BLG-RRLYR-10728 & {\bf 0.4816758(3)} & 0.3553088(5) & 0.7377 &*179.9 & *54.2 & \\
OGLE-BLG-RRLYR-11311 & 0.468425(1) & {\bf 0.3453771(4)}  & 0.7373 & *50.3 & *91.5 & \\
OGLE-BLG-RRLYR-13442 & {\bf 0.486091(1)}  & 0.354019(1) & 0.7283 & *115.3 & *44.3 & mode switch\\
OGLE-BLG-RRLYR-14915 & {\bf 0.4333758(2)} & 0.3170019(2) & 0.7315 & *99.7 & *64.1 &\\ 
OGLE-BLG-RRLYR-22356 & {\bf 0.4426324(4)} & 0.323662(1) & 0.7312 &*147.5 &  21.9 & \\ 
OGLE-BLG-RRLYR-30986 & 0.494275(1)  & {\bf 0.3630309(8)} & 0.7345 &  63.8 &  73.6 & \\ 
OGLE-BLG-RRLYR-32462 & 0.4357845(7) & {\bf 0.3222565(3)} & 0.7395 & *67.6 & *82.7 & tidal stream \\ 
\hline
\end{tabular}
\end{table*}

In a few RRd stars Blazhko-type modulation is apparent already from the investigation of the time series alone. Three examples are presented in Fig.~\ref{fig.ts}. Large modulation on a long time-scale is well visible. 

Periodic modulation of a pulsation mode manifests in the frequency spectrum as equally spaced multiplets around main frequency and its harmonics. The inverse of frequency separation between multiplet components corresponds to modulation period. Multiplets (i.e. quintuplets, septuplets, etc.) are detected in the space observations of the Blazhko RRab stars \citep[e.g.][]{chadid10,gugg} and in the top-quality ground-based observations dedicated to study of the Blazhko phenomenon \citep[e.g.][]{jurcsik_mwlyr}. In a typical ground-based data from massive sky surveys, such as OGLE or MACHO, modulation manifests as equally spaced triplets or as close doublets \citep[e.g.][]{alcock_rrab,mp03}. Following \cite{alcock_rrab} we denote these two frequency patterns as BL2 (triplets) and BL1 (doublets). Triplets are obvious signature of modulation. They are detected in four stars, only (see Tab.~\ref{tab:mod}). In the case of doublets, only one side peak is detected at $kf_0$. To interpret a close peak (doublet) as arising from modulation of a radial mode we require that it is detected not only at the primary frequency, but also at its harmonics or combination frequencies with other mode, with the same separation (and, like for all signals, its detection must be significant, with ${\rm S/N}>4$, and its frequency must be well resolved with frequency of the radial mode and other frequencies detected earlier in the analysis). In principle, doublets at $kf_0$ may correspond to a close non-radial mode and its combination frequencies with radial modes. The Blazhko (modulation) period, as defined above, is then a beating period. The more likely explanation however, is that doublets are incomplete triplets with one component of low amplitude, hidden in the noise. Because of noise present in the data, the equally spaced, but strongly asymmetric triplet may appear as doublet. Indeed, the triplets we detect in four stars may be strongly asymmetric. To quantify the asymmetry we use $Q$ parameter as defined by \cite{alcock_rrab}, $Q=(A_+-A_-)/(A_++A_-)$, where $A_{+/-}$ refers to the amplitude of the higher/lower frequency triplet component. For the 5 triplet structures reported in Tab.~\ref{tab:mod} (two triplets are detected in OGLE-BLG-RRLYR-07393) we find the following values of asymmetry parameter: $-0.36$ (OGLE-BLG-RRLYR-05762), $-0.83$ and $0.32$ (OGLE-BLG-RRLYR-07393, modulation with $\pbj$ and $\pbd$), $-0.1$ (OGLE-BLG-RRLYR-00951) and $-0.57$ (OGLE-BLG-RRLYR-02862). Strong asymmetry is not an exception. We also note that for Blazhko RRab stars \cite{alcock_rrab} analysed the distribution of side peaks' amplitudes and showed that there seems to be a continuous transition between the stars with equidistant triplet in the frequency spectrum and those showing only doublet. In a few cases the data are good enough to directly show the modulation of radial modes, their amplitudes and periods, and of mean stellar brightness, with time-dependent Fourier analysis, both for BL2 and BL1 stars. 

In the following, doublets detected in the frequency spectrum are interpreted as arising from modulation (see also Discussion in Section~\ref{ssec:dis_modp}).

In Tab.~\ref{tab:mod} we collected information about properties of the modulation detected in the analysed stars. In the following, modulation frequencies are denoted as $f_{{\rm B}i}$, and corresponding modulation periods as $P_{{\rm B}i}=1/f_{{\rm B}i}$. In each row of Tab.~\ref{tab:mod} data for one modulation period are presented. In Tab.~\ref{tab:mod} we provide the amplitude and frequency of the highest side peak at respective radial mode frequency and the associated frequency pattern (`BL2' for triplets or `BL1$+$'/`BL1$-$' for side peak on higher/lower frequency side of the radial mode). In the remarks section we provide some additional information on the frequency spectrum: `a' -- indicates that modulation peaks appear also at radial mode combination frequencies, i.e. at $kf_0\pm lf_1\pm f_{{\rm B}i}$; `b' indicates that peak is detected in the low frequency range, at $f=f_{{\rm B}i}$; `s' indicates that sub-harmonic of the modulation frequency, i.e. a peak at $f_{0/1}\pm 0.5f_{{\rm B}i}$ was detected. 

From one up to three modulation periods are present in each star. In seven stars we detected a single modulation period. Both modes may be modulated with the same period, but in a few cases each of the modes is modulated with its own period. 

In the following subsections each of the stars is analysed in detail. In Section~\ref{sec:discussion} the emerging picture of the Blazhko modulation in RRd stars is discussed in detail.

\begin{table*}
\centering
\caption{Modulation properties of the investigated RRd stars. In the consecutive columns we provide: the modulation period, the frequency of the highest modulation peak, its amplitude and structure of the modulation spectrum, for the fundamental and then for the first overtone modes. Asterisk at amplitude value indicates that corresponding peak is non-coherent. Remarks: `a' -- side peaks are also detected at frequency combinations of radial modes, $kf_0\pm lf_1$; `b' -- a peak at $f=f_{{\rm B}i}$ is detected at low frequencies; `s' -- sub-harmonic of the modulation frequency, i.e. a peak at $f_{0/1}\pm 0.5f_{{\rm B}i}$ is detected.}
\label{tab:mod}
\begin{tabular}{lrrrlrrll}
\hline
     &            &\multicolumn{3}{c}{F-mode}    &\multicolumn{3}{c}{1O-mode}\\
star & $P_{\rm B}$\thinspace [d] & freq. & $A$\thinspace [mmag] & struct. & freq. & $A$\thinspace [mmag] & struct.  & remarks\\
\hline
OGLE-BLG-RRLYR-00951 & 142.7(3)  &             &      &        & $f_1-\fbj$ & 5.5  & BL2    & a\\
\hline     
OGLE-BLG-RRLYR-02530 & 469(3)   & $f_0+\fbj$  & 22.9  & BL1$+$ & $f_1+\fbj$  & 18.7 & BL1$+$ & a\\
\hline
OGLE-BLG-RRLYR-02862 & 173.6(5) & $f_0-\fbj$ &  28.6  & BL2     & $f_1-\fbj$  & 10.3 & BL1$-$ & a\\
\hline
OGLE-BLG-RRLYR-04598 & 281(1)    & $f_0-\fbj$  & 12.9 & BL1$-$ &            &      &        & a \\ 
                     & 85.6(2)   & $f_0-\fbd$  & 10.8 & BL1$-$ &            &      &        & a \\
\hline 
OGLE-BLG-RRLYR-05762 & 97.95(2)  & $f_0-\fbj$ &*53.0 & BL2    & $f_1-\fbj$ & 5.6 & BL1$-$ & a, b\\
                     & 23.000(4) & $f_0+\fbd$ &  7.1 & BL1$+$ &            &     &        & a, b \\
                     & 81.12(8)  &            &      &        & $f_1-\fbt$ & 7.3 & BL1$-$ & a\\
\hline
OGLE-BLG-RRLYR-06283 & 332.8(3)  & $f_0-\fbj$ & *37.0 & BL1$-$ & &       &   & a, b\\
\hline
OGLE-BLG-RRLYR-07393 & 209.8(2)  & $f_0+\fbj$ &  3.8 & BL1$+$ & $f_1-\fbj$ & 50.8 & BL2 & a, b\\
                     & 322.9(4)  & $f_0+\fbd$ & *34.4 & BL1$+$ & $f_1+\fbd$ & 21.5 & BL2 & a\\
\hline
OGLE-BLG-RRLYR-09284 &  41.78(2) & $f_0+\fbj$ & *6.4 & BL1$+$ & & & & a\\   
                     &  40.76(3) & $f_0-\fbd$ &  3.8 & BL1$-$ & & & & a\\ 
                     &  47.69(5) & $f_0+\fbt$ & *2.8 & BL1$+$ & & & & a\\
\hline
OGLE-BLG-RRLYR-10728 & 98.71(8)  & $f_0+\fbj$  & 26.7 & BL1$+$ & $f_1-\fbj$ &  8.6 & BL1$-$  & a, b\\
                     & 313(1)    & $f_0-\fbd$  &  8.9 & BL1$-$ &            &      &        & a \\
\hline
OGLE-BLG-RRLYR-11311 & 49.13(2)  &             &      &        & $f_1-\fbj$ &*32.4 & BL1$-$ & a\\ 
                     & 87.1(2)   & $f_0+\fbd$  & 13.8 & BL1$+$ &            &      &        & s  \\
\hline
OGLE-BLG-RRLYR-13442 & 179.1(9)  & $f_0+\fbj$   & *14.0 & BL1$+$ &            &      &        &  \\
\hline
OGLE-BLG-RRLYR-14915 & 62.89(2)  &            &      &        & $f_1-\fbj$ & *41.2 & BL1$-$ & a, b, s\\
\hline
OGLE-BLG-RRLYR-22356 & 255.5(5) &  $f_0-\fbj$ & 40.1 & BL1$-$ &            &      &        & a, b\\
                     & 51.59(6)  &             &      &        & $f_1+\fbd$ &*15.0 & BL1$+$ & a\\
                     & 32.46(4)  &  $f_0-\fbt$ &  7.4 & BL1$-$ &            &      &        & \\
\hline 
OGLE-BLG-RRLYR-30986 & 111.4(1)  & $f_0-\fbj$ & 51.0 & BL1$-$ & $f_1-\fbj$ & 14.4 & BL1$-$ & a\\   
\hline
OGLE-BLG-RRLYR-32462 & 190.1(3)  &            &      &        & $f_1+\fbj$ &*31.4 & BL1$+$ & a\\
                     & 125.6(2)  & $f_0+\fbd$ & 23.5 & BL1$+$ & $f_1-\fbd$ &  8.8 & BL1$-$ & a\\
\hline
\end{tabular}
\end{table*}

\section{Analysis of individual stars}\label{sec:individual}

Stars are discussed in a somewhat arbitrary order. First the most interesting cases are presented, for which data allows the detailed analysis. For all discussed stars light curve solutions of OGLE-IV data are collected in Tables in the Appendix available in the on-line version of the article. Tables contain frequency identification, frequency value, amplitude and phase, all with standard errors. For a sample, see Table~\ref{tab:appsample}.

\begin{table*}
\centering
\caption{Sample table with light curve solution for OGLE-BLG-RRLYR-00951. Full table and tables for all analysed stars are in the on-line only Appendix.}
\label{tab:appsample}
\begin{tabular}{lr@{.}lrrrrr}
freq. id & \multicolumn{2}{c}{$f$\thinspace [d$^{-1}$]}& $A$\thinspace [mmag] & $\sigma$ & $\phi$ [rad] & $\sigma$ & remarks\\ 
\hline  
 $\fbj$         &  0&007010(15)  &      &      &        &      & bl\\  
 $f_1-f_0-\fbj$ &  0&6718715     &  4.3 & 0.4  &  2.57  & 0.59 &\\
 $f_1-f_0$      &  0&6788813     &  6.0 & 0.4  &  2.18  & 0.13 &\\
 $f_0$          &  1&8360933(18) &*73.4 & 0.4  &  5.64  & 0.07 &\\  
 $f_1-\fbj$     &  2&5079647     &  5.5 & 0.4  &  3.23  & 0.59 &\\
 $f_1$          &  2&5149745(21) &*58.0 & 0.4  &  1.11  & 0.08 &\\
\ldots          & \ldots&         &\ldots&\dots&\ldots&\ldots &\\
\hline
\end{tabular}
\end{table*}

\subsection{OGLE-BLG-RRLYR-05762}

\begin{figure}
\centering
\resizebox{\hsize}{!}{\includegraphics{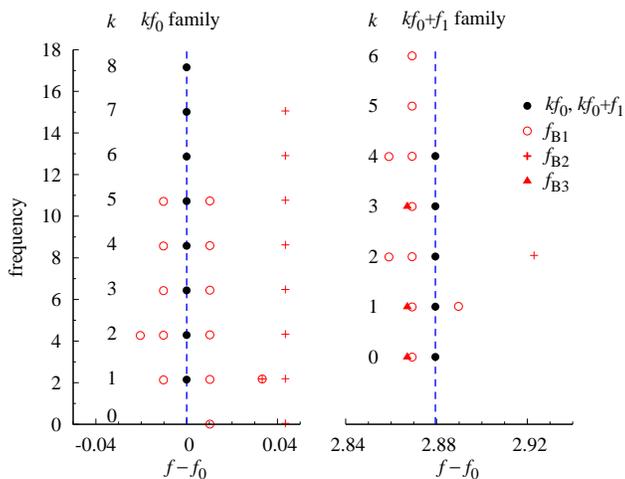}}
\caption{{\it Echelle} diagram of peaks detected at $kf_0$ and $kf_0+f_1$ for OGLE-BLG-RRLYR-05762. Side peaks with different separations are marked with different symbols.}
\label{fig.05762ech}
\end{figure}

\begin{figure*}
\centering
\resizebox{\hsize}{!}{\includegraphics{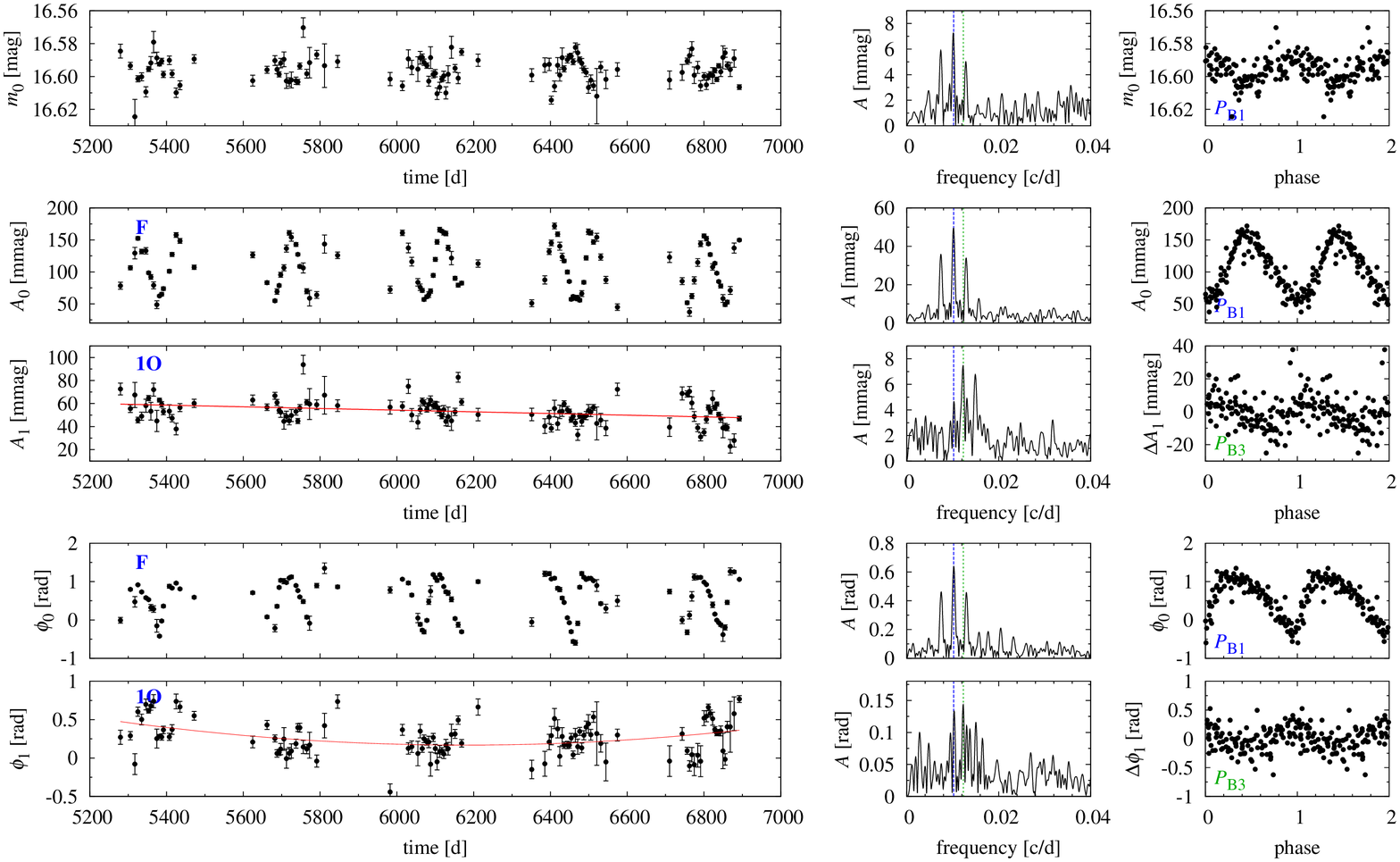}}
\caption{Time-dependent Fourier analysis for OGLE-BLG-RRLYR-05762. Left-most panels show the time variation of the mean stellar brightness, radial mode amplitudes and phases. Panels in the middle column show the frequency spectrum of the data displayed in the left column. The dashed and dotted lines in the frequency spectra mark \fbj\ and \fbt, respectively. In the right-most panels (phased modulation curves) data were phased either with \pbj ($A_0$, F-mode data) or with \pbt\ (1O data), resulting from frequency analysis. For clarity, error bars are not plotted in the phase plots (right-most panels).}
\label{fig.05762subsets}
\end{figure*}

This is one of the stars for which data are most numerous and densely sampled ($10\,359$ data points, $\rho=814$ in OGLE-IV). Pulsation in the fundamental mode dominates (Tab.~\ref{tab:dm}). The frequency spectrum is very rich. We identify three modulation periods -- Tab.~\ref{tab:mod}. The strongest modulation has period of nearly $\pbj\approx 98$\thinspace days and affects mostly the fundamental mode ($53$\thinspace mmag at $f_0-\fbj$). A family of triplets is well visible at all harmonics till $5f_0$ (Fig.~\ref{fig.05762ech}). In addition, a quintuplet component of lower frequency is present at $2f_0$. A strong peak at $f=\fbj$ is also detected ($6.7$\thinspace mmag). First overtone is also affected by this modulation, very weakly, however. We only detect one significant (${\rm S/N}=8.9$) side peak at $f_1-\fbj$ and its amplitude is ten times smaller than in the case of the fundamental mode ($5.6$\thinspace mmag). In addition to the triplets with $\fbj$ separation we find two families of doublets with $\fbd$ and $\fbt$ separations. The side peaks with $\fbd$ separation ($\pbd=23$\thinspace days) appear on the higher frequency side of $f_0$, i.e. at $kf_0+\fbd$ and are detected at all harmonics up to $7f_0$ (Fig.~\ref{fig.05762ech}). The amplitude of the highest side peak (at $2f_0+\fbd$) is $8.3$\thinspace mmag. Combination frequencies are also accompanied by \fbd\ but we do not detect such peaks at the first overtone frequency and at its harmonics ($kf_1$). The side peaks with \fbt\ separation ($\pbt=81.1$\thinspace days) are detected on lower frequency side of $f_1$, $2f_1$ and at combination frequencies with $f_0$, but not at $kf_0$. In Fig.~\ref{fig.05762ech} we show all detected peaks at $kf_0$ and at combination frequencies $kf_0+f_1$ in the form of {\it echelle} diagram, i.e. frequency is plotted vs. $f-f_0$ \citep{gugg}.

As data sampling is very dense we conducted the time-dependent Fourier analysis ($\Delta t\approx 5$\thinspace d) which confirms the picture emerging from the frequency analysis. Results are collected in Fig.~\ref{fig.05762subsets}. The left panels show, from top to bottom: time variation of the mean stellar brightness, $m_0$, variation of the amplitude of the fundamental and of the first overtone modes ($A_0$ and $A_1$), and variation of their phases ($\phi_0$ and $\phi_1$). Middle panels show the Fourier transform of the time series' presented in the left panels, and right-most panels are phased modulation curves. We note that in the top left panel (mean-stellar brightness) a possible season-to-season differences in the photometric zero-point will also be revealed, but we believe these are small.

Modulation of the fundamental mode with \pbj, both amplitude and phase modulation, is obvious. Suspected modulation of the fundamental mode with $\pbd$ is masked by a much larger modulation with \pbj. $\pbd$ is also too short to be well resolved in the time-dependent Fourier analysis. In the case of first overtone, modulation with \pbt\ is dominant.  We observe the following correlations: ({\it i}) the mean stellar brightness is modulated with $\pbj$ and is anti-correlated with the amplitude of the fundamental mode: lower the pulsation amplitude, brighter the star; ({\it ii}) Both amplitude and phase of the fundamental mode are modulated with $\pbj$; ({\it iii}) The lowest amplitude of the fundamental mode corresponds to the longest value of its period; the following period decrease is faster than the increase of pulsation amplitude (see eq.~\ref{eq:pc}); ({\it iv}) First overtone vary on a long time-scale: its mean amplitude decreases, so is its period (at an estimated rate of $1.4\times 10^{-5}$\thinspace d per $1000$\thinspace d). These long-term trends (solid lines in the left-most panels of Fig.~\ref{fig.05762subsets}) are removed before analysing short-term modulation; ({\it v}) Both amplitude and phase of the first overtone are modulated with $\pbt$; ({\it vi}) Because of low modulation amplitude, the phase relation between amplitude and phase modulation for the first overtone is not as clearly marked as in the case of the fundamental mode.

\begin{figure}
\centering
\resizebox{\hsize}{!}{\includegraphics{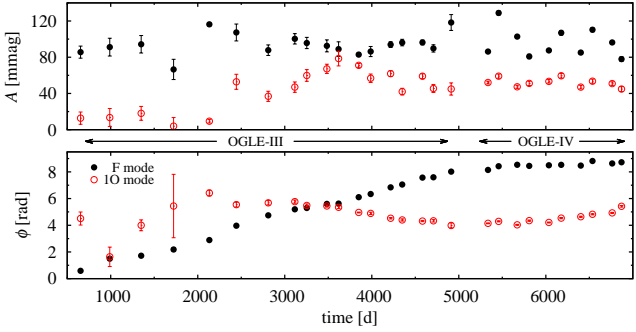}}
\caption{Time-dependent Fourier analysis for OGLE-BLG-RRLYR-05762 ($\Delta t\approx 150$\thinspace d).}
\label{fig.05762tdfd}
\end{figure}

OGLE-III data for OGLE-BLG-RRLYR-05762 exist but the observation cadence is much lower ($\rho=115$). Only four significant side peaks are detected in the frequency spectrum, to be compared with $43$ detected in OGLE-IV data. One of the side peaks appears nearly exactly at $f_0+\fbd$  just as in the case of OGLE-IV data. Other side peaks appear on lower frequency side of $f_0$ and of $f_1$. The inverse of frequency separation corresponds to modulation period of $91$\thinspace days -- in between $\pbj$ and $\pbt$ detected in OGLE-IV data. We note a significant power excess in large frequency range around both $f_0$ and $f_1$ -- OGLE-III data are not good enough to analyse the short-term modulation in detail, but indicate that it occurs. It is possible however to analyse long-term variation of radial modes. In Fig.~\ref{fig.05762tdfd} we present the results of time-dependent-Fourier analysis with $\Delta t\approx 150$\thinspace d, which effectively averages the shorter period modulations. Both modes are non-stationary on a long time-scale. We note a very low amplitude of the first overtone at the first seasons of OGLE-III observations and pronounced phase (period, eq.~\ref{eq:pc}) change of both modes. At $2500\,{\rm d}<t<4500\,{\rm d}$ the amplitudes of both modes are anti-correlated.

\subsection{OGLE-BLG-RRLYR-06283}

It is the second star in the analysed sample with the largest number of points ($10\,310$ data points, $\rho=810$ in OGLE-IV). Pulsation in the first overtone is dominant (Tab.~\ref{tab:dm}). Its frequency spectrum is rather simple. Except $f_0$, $f_1$ and their combinations we see side peaks on the lower frequency side of $f_0$, $2f_0$ and $3f_0$, and at detected combination frequencies with $f_1$ (doublets). The highest side peak at $f_0-\fbj$ has an amplitude of $37$\thinspace mmag and is non-coherent. After prewhitening a noticeable power excess is detected at $f_0$. One of the peaks closet to $f_0$ may correspond to the high frequency triplet component, but it is weak and non-coherent.

\begin{figure*}
\centering
\resizebox{\hsize}{!}{\includegraphics{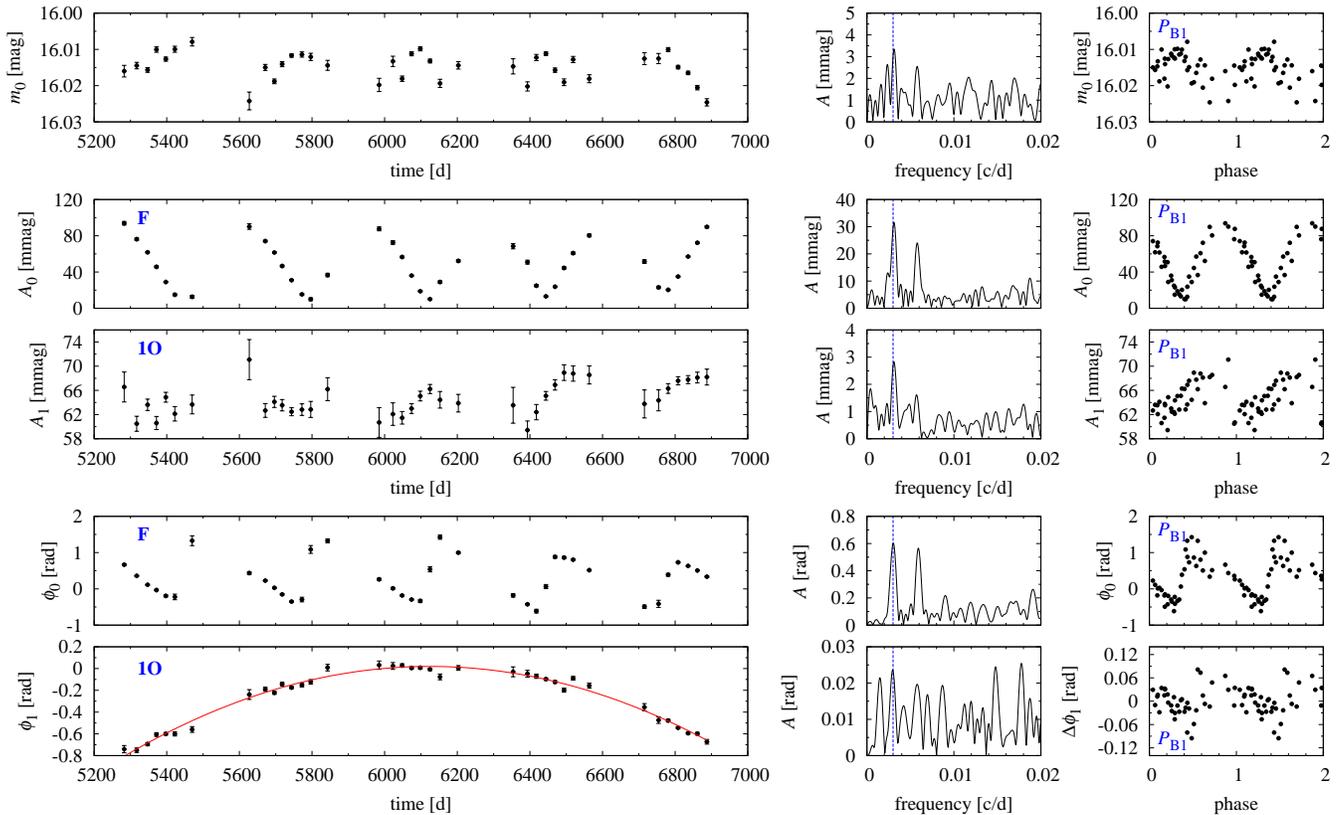}}
\caption{Time-dependent Fourier analysis for OGLE-BLG-RRLYR-06283. The dashed lines in the frequency spectra mark \fbj. In the rightmost panels all data were phased with \pbj\ resulting from frequency analysis. For clarity, error bars are not plotted in the phase plots (right-most panels).}
\label{fig.06283subsets}
\end{figure*}

The low frequency signal at $f=\fbj$ is firmly (${\rm S/N}=6.5$) detected. The signal at $kf_1$ is non-coherent and cannot be removed with a single sine wave. We detect bunches of closely spaced peaks at $kf_1$. Such structure is a signature of strong period and/or amplitude change of the first overtone which may hide the possible modulation structure. This is confirmed with the time-dependent Fourier analysis ($\Delta t\approx 25$\thinspace d), results of which are presented in Fig.~\ref{fig.06283subsets}. We observe: ({\it i}) A weak signature of modulation of mean stellar brightness with $\pbj$. The mean stellar brightness is anti-correlated with amplitude of the fundamental mode: lower the pulsation amplitude, brighter the star; ({\it ii}) Both amplitude and phase of the fundamental mode are modulated with $\pbj$; amplitude modulation is more symmetric than phase modulation; ({\it iii}) At the phase slightly preceding the minimum pulsation amplitude of the fundamental mode its period is longest. The following period decrease is very fast, lasts roughly 1/3 of the modulation cycle, and occurs at a phase of small amplitude of the fundamental mode; ({\it iv}) Period of the first overtone vary on a long time-scale. It increases at an estimated rate of $5\times 10^{-5}$\thinspace d per $1000$ days. The parabolic trend was removed from phase variation before we searched for short-term modulation of the first overtone; ({\it v}) We find a weak signature of the modulation of the first overtone with $\pbj$. The amplitude modulation is best visible, but weak ($2.8$\thinspace mmag as determined from the frequency spectrum). Phase modulation may be masked by the use of approximate model for long-term variation (parabolic trend); ({\it vi}) Phase relation between modulation of the fundamental and first overtone modes is not clear: first, the phase of maximum amplitude of the fundamental mode is weakly sampled, second, modulation of the first overtone is weak and clearly non-stationary (left panel in Fig.~\ref{fig.06283subsets}).

OGLE-III data for the star are insufficient for detailed frequency analysis ($\rho=118$). Only one significant side peak is detected in the frequency spectrum. The merged data set allows to study the long term variation of radial modes. This analysis confirms a fast and linear change of the first overtone period. The phase of the fundamental mode is stable. We only note a $\sim 20$ per cent lower mean amplitude of the fundamental mode within OGLE-III data.

\subsection{OGLE-BLG-RRLYR-07393}\label{sec:07393}

In this star pulsation in the first overtone dominates (Tab.~\ref{tab:dm}). We clearly detect modulation with two periods, $\pbj=209.8$\thinspace d and $\pbd=322.9$\thinspace d (Tab.~\ref{tab:mod}). Both modes are affected by these two modulations but to a different extent. The modulation is strongest for the first overtone -- two equidistant triplets, one with $\fbj$ separation, the other with $\fbd$ separation, are detected at $f_1$. The highest side peak, at $f_1-\fbj$ has an amplitude of $50.8$\thinspace mmag and the highest side peak connected with $\pbd$\ has an amplitude of $21.5$\thinspace mmag (at $f_1+\fbd$). In the case of fundamental mode, modulation with \pbd\ is dominant ($34.4$\thinspace mmag at $f_0+\fbd$), while modulation with \pbj\ is very weak ($3.8$\thinspace mmag at $f_0+\fbj$). We note that $\pbj/\pbd\approx 2\!\!:\!\!3$. The beat frequency is $f_{\rm beat}=\fbj-\fbd\approx 0.00167$\thinspace c/d which corresponds to beat period of $\approx 600$\thinspace days. Nearly $3$ beat periods are covered in the OGLE-IV data. The beating is well visible in the envelope of the full light curve solution which we schematically plotted with the dashed line in Fig.~\ref{fig.ts}. 

Strong modulation of the first overtone is reflected in rich modulation pattern at $f_1$. We detect triplet $f_1\pm\fbj$, incomplete quintuplet, i.e. peaks at $f_1\pm\fbd$ and at $f_1-2\fbd$, and triplet with beat frequency, i.e. $f_1\pm f_{\rm beat}$. Exactly the same modulation pattern is detected at combination frequency, $f_0+f_1$. After prewhitening with these frequencies, an excess power is still detected at $f_1$. The highest peak is at location unresolved with $f_1$ (${\rm S/N}=7.9$). The second highest peak is located close, but not exactly at $f_1-\fbj-\fbd$. Prewhitening with this frequency does not remove the signal. The excess power at $f_1$ is a signature of non-stationary nature of the first overtone and its modulation.

In the low frequency range we clearly (${\rm S/N}=11$) detect a peak at $f=\fbj$. In addition there is a peak at $2\fbd$  (${\rm S/N}=4.8$) and likely at $\fbd$ (${\rm S/N}=5.2$) but the latter detection is uncertain. At even lower frequency there is a peak of slightly higher amplitude, which may be a signature of weak long-term trend. These two peaks are 1-year aliases of each other.  

After prewhitening the data with $f_0$, $f_1$, their combinations and modulation components, additional signal remains in the data. We detect four significant peaks located at frequencies corresponding to the inverse of sideral day and its three consecutive harmonics. We treat this signal as of instrumental origin. In addition we find a significant (${\rm S/N}=4.8$) peak at $f_{\rm x}\approx 4.705$\thinspace c/d ($A_{\rm x}=2.9$\thinspace mmag). The period ratio $P_{\rm x}/P_{1}=0.6163$ indicates that $P_{\rm x}$ cannot correspond to a radial mode. This untypical period ratio is not new for RR~Lyrae stars however. It is found in both RRc and RRd stars. It corresponds to the new class of radial--non-radial RR~Lyrae stars \citep[Introduction,][]{pamsm14,netzel}. In particular, $P_{\rm x}/P_1$ perfectly fits the sequence of more than 100 such stars detected in the OGLE-III data of the Galactic bulge \citep[see Petersen diagram in fig.~7 in][]{netzel}. We also find a signature of modulation of non-radial mode. At $f_{\rm x}-\fbd$ we detect a weak (${\rm S/N}=4.1$, $2.5$\thinspace mmag) peak, which indicates that non-radial mode is modulated with the same period as the two radial modes.

\begin{figure*}
\centering
\resizebox{\hsize}{!}{\includegraphics{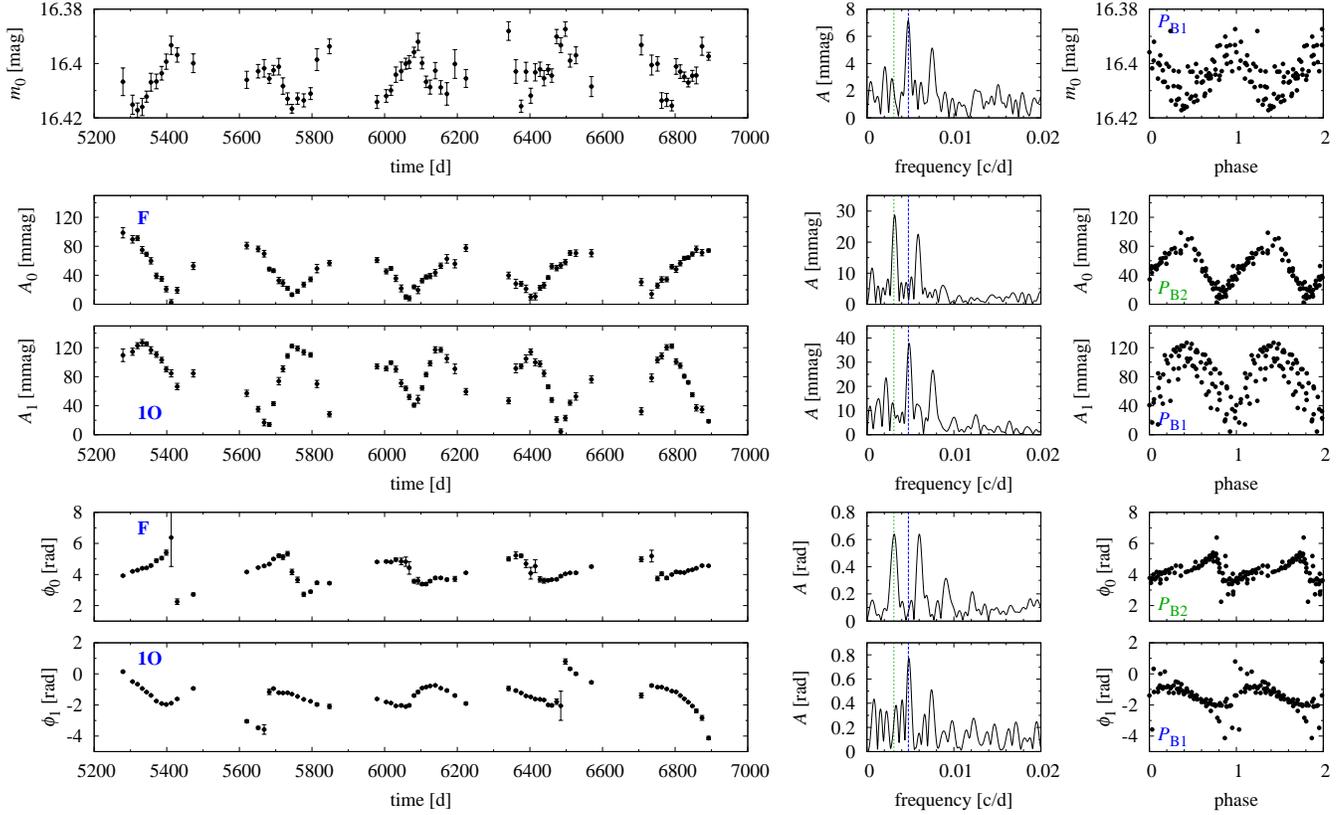}}
\caption{Time-dependent Fourier analysis for OGLE-BLG-RRLYR-07393. The dashed and dotted lines in the frequency spectra mark \fbj\ and \fbd, respectively. In the right-most panels data were phased either with \pbj ($A_0$, 1O-mode data) or with \pbd\ (F-mode data), resulting from the frequency analysis. For clarity, error bars are not plotted in the phase plots (right-most panels).}
\label{fig.07393subsets}
\end{figure*}

In Fig.~\ref{fig.07393subsets} we present the results of the time-dependent Fourier analysis ($\Delta t\approx 11$\thinspace d), which confirms the picture emerging from the frequency analysis. In the case of the fundamental mode, modulation with \pbd\ is apparent and dominant. The phased modulation curve is sharp, with little scatter, as modulation amplitude with \pbj\ is ten times smaller. This is not the case for the first overtone.  The modulation of the first overtone with \pbj\ is well visible. The phased modulation curve has large scatter however, as modulation with both periods is strong (Tab.~\ref{tab:mod}). The behavior of stellar mean brightness is clearly anti-correlated with the amplitude of the first overtone mode. When amplitude of the first overtone drops, the mean brightness of the star increases. In the phased modulation curve (top right panel in Fig.~\ref{fig.07393subsets}) the same scatter as in the case of first overtone is observed. As modulations dominant for the fundamental mode and for the first overtone have different periods, there is no obvious correlation between the amplitudes of the two modes. Sometimes they vary in phase, sometimes the amplitudes are anti-correlated -- a pattern that repeats with a beat period.

Of particular interest is behaviour of mode phases and their correlation with mode amplitudes. Phases (pulsation periods) are modulated with the period dominant for a given mode (\pbj\ for the first overtone, \pbd\ for the fundamental mode). We recall that in case of OGLE-BLG-RRLYR-05762 and OGLE-BLG-RRLYR-06283 this modulation was smooth and continuous. This is not the case for OGLE-BLG-RRLYR-07393, however. We observe phase jumps when the corresponding mode amplitude is lowest (left-most panels in Fig.~\ref{fig.07393subsets}). Unlike in the two mentioned stars the amplitudes of the fundamental and of the first overtone modes are very low (nearly vanish) from time to time. The apparent discontinuity in the pulsation phase is then detected (see also Discussion in Sec.~\ref{ssec:dis_modp}).

\begin{figure*}
\centering
\resizebox{\hsize}{!}{\includegraphics{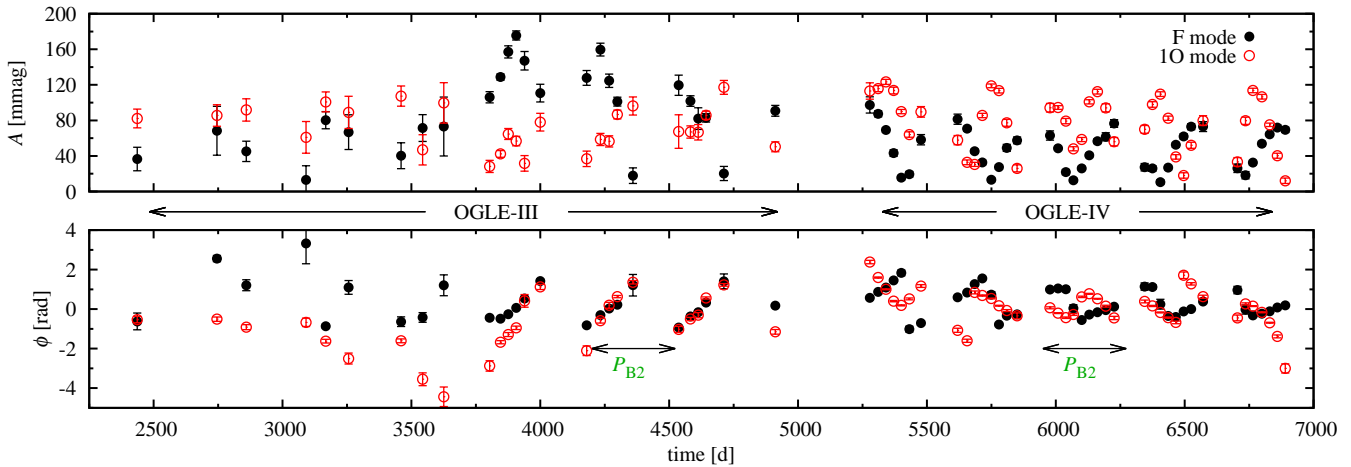}}
\caption{Time-dependent Fourier analysis for OGLE-III and OGLE-IV data for OGLE-BLG-RRLYR-07393.}
\label{fig.07393tdfd}
\end{figure*}

The sampling of OGLE-III data for this star is lower than in OGLE-IV ($\rho=66$ vs. $\rho=322$ in OGLE-IV). Their independent analysis yields a somewhat different frequency spectrum. The resulting modulation properties are summarised in Tab.~\ref{tab:mod2}. Two modulation periods are apparent, which we denote with `$'$'. $\pbd'\approx\pbd$, but $\pbj'=180$\thinspace days and is significantly shorter than $\pbj$ (Tab.~\ref{tab:mod}). We observe triplet structures at the fundamental mode frequency rather than at the first overtone frequency. In the case of the first overtone and $\pbd'$ a triplet is also likely, but higher frequency component overlaps with 1-year alias of unresolved signal at $f_1$. For both modes modulation with $\pbd'\approx\pbd$ is dominant within OGLE-III. This is visible in the time-dependent Fourier analysis for the merged data, results of which are presented in Fig.~\ref{fig.07393tdfd}. The last three seasons of OGLE-III data, $3750\,{\rm d}<t<4750\,{\rm d}$, are densely sampled and allow to see the modulation. Phase plots provide the most clear view. It is visible that the dominant period of modulation of the first overtone's phase is $\pbd'=\pbd$ at the end of OGLE-III observations, and a shorter period ($\pbj$) during OGLE-IV observations (see also Fig.~\ref{fig.07393subsets}). Modulation of the fundamental mode is always dominated by $\pbd$. Its mean amplitude vary on a long time-scale with the event of significantly larger amplitudes at $3750\,{\rm d}<t<4500\,{\rm d}$.

\begin{table*}
\centering
\caption{The same as Tab.~\ref{tab:mod}, but for OGLE-BLG-07393 and OGLE-III data only.}
\label{tab:mod2}
\begin{tabular}{lrrrlrrll}
\hline
     &            &\multicolumn{3}{c}{F-mode}    &\multicolumn{3}{c}{1O-mode}\\
star & $P_{\rm B}'$\thinspace [d] & freq. & $A$\thinspace [mmag] & struct. & freq. &  $A$\thinspace [mmag] & struct.  & remarks\\
\hline 
OGLE-BLG-RRLYR-07393 & 180.4(4)  & $f_0+\fbj'$ &  *23.2 & BL2 & $f_1-\fbj'$ & *18.3 & BL1$-$ & \\
                     & 322.4(1.1)  & $f_0-\fbd'$ & *39.5 & BL2 & $f_1-\fbd'$ & *38.8 & BL1$-$ & b \\
\hline
\end{tabular}
\end{table*}

Based on the data we have we cannot judge what has happened with the shorter modulation period. We lack the crucial data in between OGLE-III and OGLE-IV, also lower cadence of OGLE-III observations data does not help to resolve the issue. We also note that all modulation side peaks are non-coherent (Tab.~\ref{tab:mod2}). The modulation with the shorter period, as well as both fundamental and first overtone modes, are non-stationary on a longer time-scales.

\subsection{OGLE-BLG-RRLYR-14915}\label{sec:14915}

For this star fundamental mode pulsation dominates (Tab.~\ref{tab:dm}). The highest side peak, with amplitude of $41.2$\thinspace mmag, is detected at the first overtone frequency ($f_1-\fbj$; $\pbj=62.89$\thinspace d). A side peak with the same separation also appears at $2f_1$ and at all detected combination frequencies involving $f_0$ and $f_1$ (doublets). A low frequency peak at $f=\fbj$ is well visible (${\rm S/N}=3.7$). At $f_1$ we also detect a side peak exactly at $f_1-\fbj/2$. We postpone discussion of its origin till the end of this subsection.

Interestingly, side peak with different separation, $\fbd$ ($\pbd=140.2$\thinspace d), is detected only at the combination frequencies: $f_0+f_1+\fbd,\ 2f_0+f_1+\fbd,\ 2f_0+2f_1+\fbd$, and at $f_1-f_0-\fbj-\fbd$, but not directly at $f_0$ or at $f_1$. These peaks (at combination frequencies) have low amplitudes. The highest, at $f_0+f_1+\fbd$, has an amplitude of $5.7$\thinspace mmag (${\rm S/N}=6.0$). As no side peak with $\fbd$ separation is detected directly at $f_0$ or at $f_1$, we cannot claim the modulation of these modes with $\pbd$ and therefore, no information is provided in Tab.~\ref{tab:mod}. We note that other interpretation of these peaks is possible, namely, the highest peak at $f_0+f_1+\fbd$ can correspond to the fourth radial overtone, $f_4$. Other peaks involving $\fbd$ are then linear combinations of $f_0$, $f_1$, $f_4$ and $\fbj$. Indeed, $P_4/P_1=0.577$ which is close to the expected period ratio of the radial modes -- see e.g. fig.~4 in \cite{netzel}. The best match with the models is obtained assuming low metallicity, but then we cannot match the $P_1/P_0$ ratio, which requires high metallicity \citep[see the Discussion in Section 5.2 and][for similar considerations for other star]{smolec_tri}. Therefore, such explanation also faces a difficulty.

\begin{figure}
\centering
\resizebox{\hsize}{!}{\includegraphics{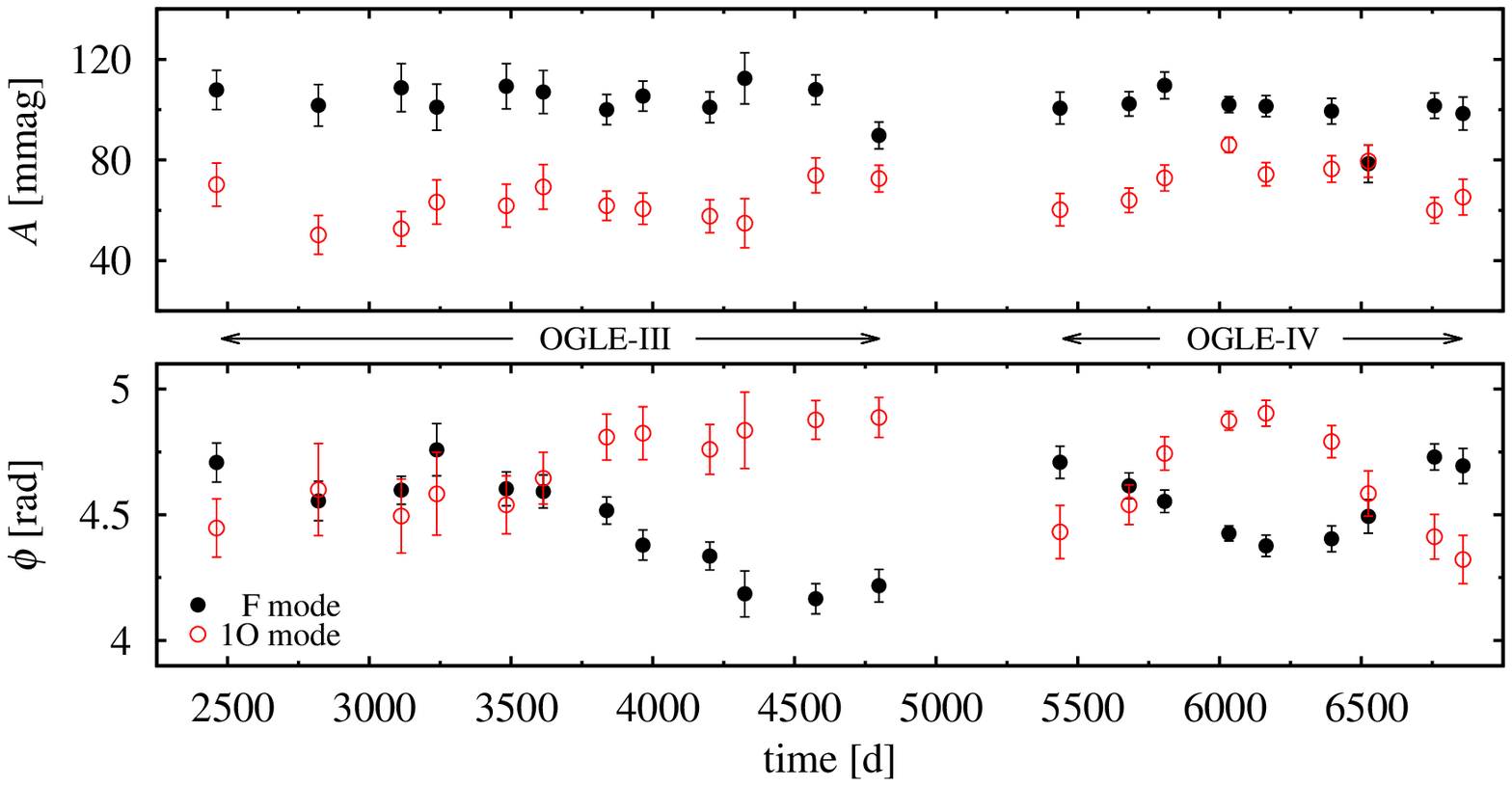}}
\caption{Time-dependent Fourier analysis for OGLE-BLG-RRLYR-14915 ($\Delta t\approx 120$\thinspace d).}
\label{fig.14915tdfd}
\end{figure}

In the frequency spectrum of residual data we observe unresolved power at both $f_0$ and $f_1$ which may increase the noise level in the transform and hide the modulation peaks. It indicates that both modes undergo period and/or amplitude change on a long time-scale. To investigate it we merged OGLE-III and OGLE-IV data and conducted a time-dependent Fourier analysis ($\Delta t\approx 120$\thinspace d), to search for a possible long-term modulation (for this star data are insufficient to investigate short-term modulation with $\pbj$ or $\pbd$; $\rho=39/87$ in OGLE-III/OGLE-IV). Results are presented in Fig.~\ref{fig.14915tdfd}. Phases of the fundamental and of the first overtone modes vary on a long time-scale and are anti-correlated. The variation is not single-periodic however; based on the data we have we cannot judge whether it is multi-periodic or irregular. The amplitudes also vary on the same time-scale and for most of the time are also anti-correlated but the effect is not as strong as in the case of pulsation phases. In the Fourier transform of the combined OGLE-III and OGLE-IV data we observe weakly resolved, and non-coherent peaks at both $f_0$ and $f_1$ in agreement with the just described picture.

Since the variation presented in Fig.~\ref{fig.14915tdfd} occurs on a long time-scale, and associated unresolved power at $f_0$ and $f_1$ increases the noise level in the transform, we decided to filter out this long-term variation using time-dependent prewhitening. With $\Delta t\approx 300$\thinspace d most of the long-term variation should be filtered out, while leaving the short-term variation (associated with \pbj\ and \pbd) unchanged. Main pulsation frequencies (and their combinations) should also be removed provided the associated phase/amplitude variation is small within a $\approx 300$\thinspace d data chunk. This is not the case however, and small residual power at $f_0$, $f_1$, $2f_1$ and $f_1-f_0$ remains. The secondary side peaks dominate the Fourier spectrum; all detected in the original, combined OGLE-III and OGLE-IV data are recovered, however no new significant detections are made.

 We recall that in OGLE-IV data, but also in OGLE-III and in the merged data we detect a peak exactly (within frequency resolution) at $f_1-\fbj/2$. This is a significant (${\rm S/N}=5.2$ in OGLE-IV, ${\rm S/N}=6.5$ in the merged data) detection with peak amplitude of $4.8$\thinspace mmag in the OGLE-IV data set. Three interpretations are possible: ({\it i}) $\fbj/2$ corresponds to basic modulation frequency and $\fbj$ is its harmonic. This is unlikely however, as peak at $f_1-\fbj/2$ is weak ($4.8$\thinspace mmag to be compared with $41.2$\thinspace mmag at $f_1-\fbj$ in the OGLE-IV) and is not detected at $2f_1$ or at combination frequencies where doublet with \fbj\ separation is present; ({\it ii}) $\fbj/2$ may correspond to the independent modulation period for first overtone mode; ({\it iii}) $\fbj/2$ is sub-harmonic of modulation frequency and thus full modulation cycle lasts $2\pbj$ with alternating modulation cycles. This, in our opinion, is the most likely explanation. In other star, OGLE-BLG-RRLYR-11311, we find a similar frequency pattern.

\subsection{OGLE-BLG-RRLYR-09284}

The star is a likely member of a tidal stream crossing the Galactic bulge \citep{ogleiv_rrl_blg}. Pulsation in the first overtone is dominant (Tab.~\ref{tab:dm}). We detect three side peaks at fundamental mode frequency, all with low amplitudes, below $7$\thinspace mmag. The two highest side peaks appear on either side of $f_0$ and at first glance resemble equally spaced triplet. This is not the case however. The separations correspond to slightly different modulation periods: $\pbj=41.78$\thinspace d ($f_0+\fbj$, $6.4$\thinspace mmag) and $\pbd=40.76$\thinspace d ($f_0-\fbd$, $3.8$\thinspace mmag). Analysis of the merged OGLE-III and OGLE-IV data confirms the asymmetry. The two frequencies $\fbj$ and $\fbd$ are well resolved in a merged data set. In addition, on the higher frequency side of $f_0$ we find additional side peak at $f_0+\fbt$ ($\pbt=47.7$\thinspace d). The doublets are also found at combination frequencies with the first overtone mode and at $2f_0$ ($2f_0+\fbj$). A significant unresolved power remains at $f_0$ after prewhitening (and also at the side peaks). 

At first overtone frequency we detect a bunch of close (weakly resolved) peaks which may correspond to long-term variability of the first overtone. They are present only at $f_1$ and not at frequency combinations with $f_0$ and have small amplitudes (all below $1.5$\thinspace mmag). As such they are not convincing signature of modulation. There is a significant single peak on the higher frequency side of $f_1$, at $f_{\rm u}$ (separation with $f_1$ is $\approx 0.075$\thinspace c/d, ${\rm S/N}=5.6$), but we do not find any combination frequency involving $f_{\rm u}$ in the spectrum. In addition we find weak peaks at $f_1+\fbj$ and $f_1-\fbj$ (${\rm S/N}$ equal to $3.4$ for both peaks). They likely correspond to the modulation present for the fundamental mode, but the detections are marginal, and hence not reported in Tab.~\ref{tab:mod}.

\begin{figure}
\centering
\resizebox{\hsize}{!}{\includegraphics{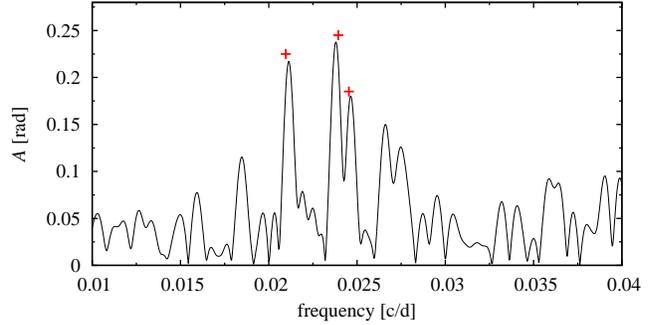}}
\caption{Frequency spectrum for fundamental mode phases extracted with time-dependent Fourier analysis for OGLE-BLG-RRLYR-09284. `+' signs mark the location of modulation frequencies derived from the analysis of OGLE-IV data (Tab.~\ref{tab:mod}).}
\label{fig.09284fsp}
\end{figure}

For this star the data have relatively good sampling ($\rho=320$) but suspected modulation is of low amplitude. Nevertheless we attempted the time-dependent Fourier analysis with short $\Delta t\approx 7$\thinspace d to capture the modulation of the fundamental mode. Only in the case of the fundamental mode's phase we find a weak evidence of its modulation with the three reported periods. In the Fourier transform of fundamental mode phases, Fig.~\ref{fig.09284fsp}, we first detect a signal at $\fbj$, and after prewhitening also with $\fbd$ and $\fbt$. 

\begin{figure}
\centering
\resizebox{\hsize}{!}{\includegraphics{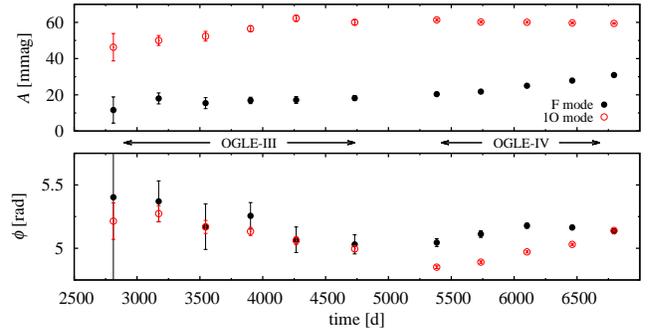}}
\caption{Time-dependent Fourier analysis for OGLE-BLG-RRLYR-09284 ($\Delta t\approx 300$\thinspace d). For better visibility the phase of first overtone was shifted vertically by $1$\thinspace rad.}
\label{fig.09284tdfd}
\end{figure}

For the first overtone signatures of modulation (only with $\pbj$) are very weak, if any. They may be masked however by long-term variation of the first overtone mode. We also expect such long-term variation for fundamental mode as at $f_0$ we detect unresolved residual power. To investigate the nature of long-term variability of the first overtone and of the fundamental mode we conducted time-dependent Fourier analysis of the merged data ($\Delta t\approx 300$\thinspace d). Results are presented in Fig.~\ref{fig.09284tdfd}. Long-term variability of both modes is well visible. The phases seem to be correlated and vary on a long time-scale. Interestingly, a significant and steady increase of the fundamental mode amplitude is well visible within the OGLE-IV data. First overtone amplitude is roughly constant then. Reverse is true during early seasons of the OGLE-III observations, but the trends are less significant, also due to larger errors caused by sparse sampling within OGLE-III ($\rho=40$).

The described long-term variability was subtracted from the data using the time-dependent prewhitening. We recover all modulation components detected in the analysis of OGLE-IV data, however do not find any additional, significant side peak.

\subsection{OGLE-BLG-RRLYR-32462}

For this star data were collected only during the OGLE-IV phase. The star is a likely member of a tidal stream crossing the Galactic bulge \citep{ogleiv_rrl_blg}. Pulsation in the first overtone is dominant and so is its modulation with $\pbj=190.1$\thinspace d (doublets on higher frequency side, the highest peak at $f_1+\fbj$ has an amplitude of $31.4$\thinspace mmag). A side peak with \fbj\ separation is not detected at $f_0$. We only detect a very weak signals at $2f_0+2\fbj$ and $2f_0+\fbj$ (${\rm S/N}$ equal to $3.4$ and $3.2$, respectively), which may indicate that also fundamental mode is modulated with $\pbj$ but firm evidence is missing. No doubt fundamental mode is modulated with a shorter period, $\pbd=125.6$\thinspace d ($23.5$\thinspace mmag at $f_0+\fbd$). The same modulation affects the first overtone ($8.8$\thinspace mmag at $f_1-\fbd$). The fundamental mode is clearly non-stationary (unresolved power at $f_0$ with ${\rm S/N}=15.3$). Unresolved power also remains at $f_1$ (${\rm S/N}=7.1$) but the signal is lower than non-coherent signal at $f_1+\fbj$ (${\rm S/N}=9.5$).

\begin{figure*}
\centering
\resizebox{\hsize}{!}{\includegraphics{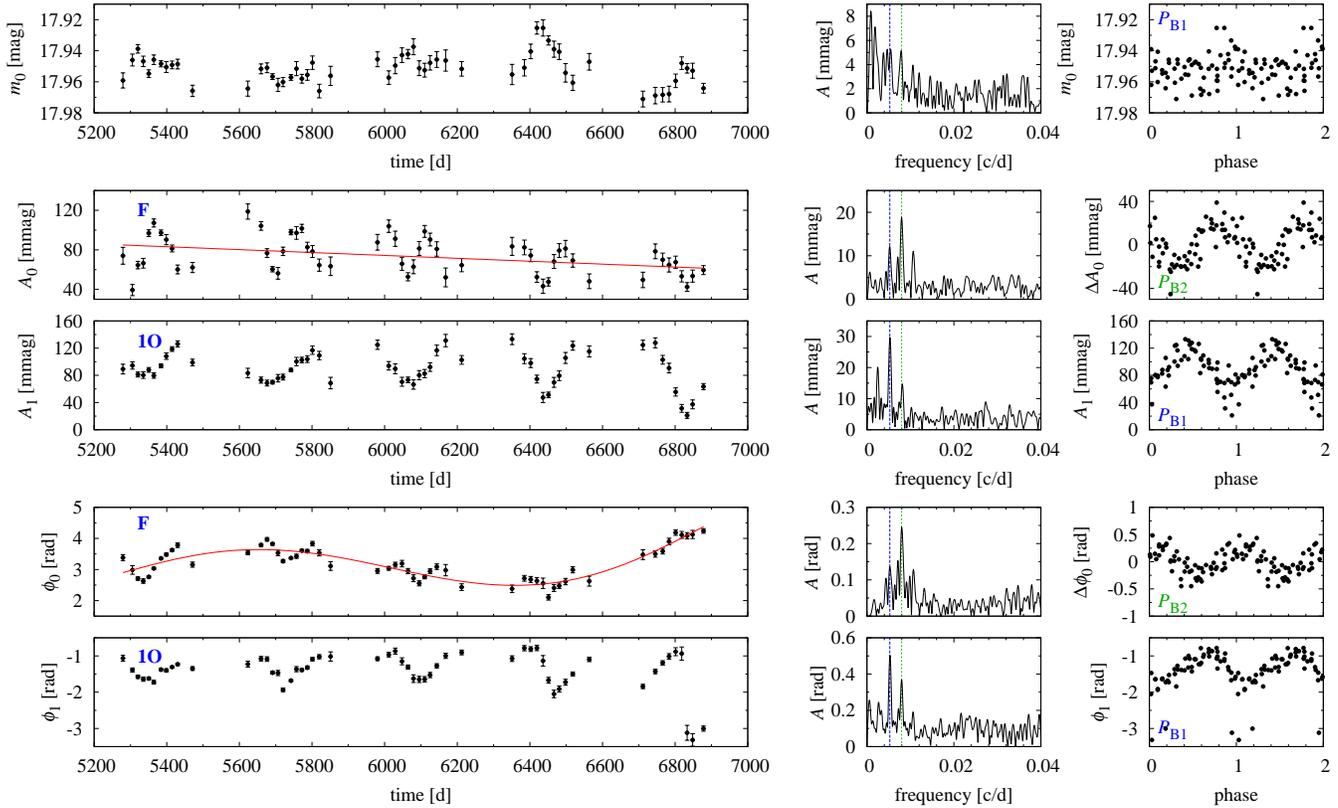}}
\caption{Time-dependent Fourier analysis for OGLE-BLG-RRLYR-32462. The dashed and dotted lines in the frequency spectra mark \fbj\ and \fbd, respectively. In the right-most panels data were phased either with \pbj ($A_0$, 1O-mode data) or with \pbd\ (F-mode data), resulting from the frequency analysis.}
\label{fig.59913subsets}
\end{figure*}

Although in the frequency spectrum we find doublets only, we conclude that both modes are modulated with the two periods. Modulation is well visible in the time-dependent Fourier analysis ($\Delta t\approx 15$\thinspace d, Fig.~\ref{fig.59913subsets}). We observe the following: ({\it i}) Fundamental mode, both amplitude and phase, is modulated with $\pbd$. The amplitude of the fundamental mode steadily decreases, its phase underwent a large and long-term variation which we modelled with fifth-degree polynomial. As a result of this long-term variation the peak at $f_0$ in the frequency spectrum of original data is non-coherent as described above; ({\it ii}) First overtone, both amplitude and phase, are modulated with $\pbj$. The amplitude of modulation clearly increases. As a result a peak at $f_1+\fbj$ is non-coherent as pointed above; ({\it iii}) When amplitude of first overtone is significantly reduced (around $t=6800$\thinspace d) a glitch in its phase variation is observed; ({\it iv}) No clear modulation of the mean stellar brightness is detected in the Fourier spectrum and in the phased data for analysed subsets. However, the correlation between $m_0$ and $A_1$ is clearly visible in the left-most panels of Fig.~\ref{fig.59913subsets}, in particular during the fourth and fifth observing seasons: mean brightness increases when amplitude of first overtone decreases and {\it vice versa}. The effect is not as clear in the earlier observing seasons as modulation amplitude of the first overtone is smaller then.

Time-dependent prewhitening does not provide additional insight into modulation properties of the star. Modulation periods are rather long and hence, it is difficult to pick $\Delta t$ such that long-term variation is eliminated, while periodic modulation remains unaffected.

\subsection{OGLE-BLG-RRLYR-22356}

\begin{figure}
\centering
\resizebox{\hsize}{!}{\includegraphics{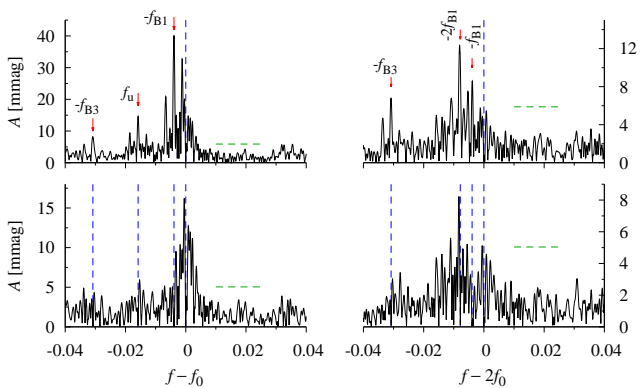}}
\caption{Prewhitening of the side peaks at $f_0$ and $2f_0$ for OGLE-BLG-RRLYR-22356. The short horizontal line segments are placed at four times the noise level in the frequency spectrum. Arrows in the top panels indicate the significant frequencies. $f_0$ and $2f_0$ are prewhitened in all panels (the right-most dashed lines).}
\label{fig.85370fsp}
\end{figure}

For this star data were collected only during the OGLE-IV phase. Pulsation in the fundamental mode dominates. On the lower frequency side of $f_0$ we detect three significant close peaks. Side peaks at  $f_0-\fbj$ and at $f_0-\fbt$ have amplitudes of $40.1$ and $7.4$\thinspace mmag, respectively (periods corresponding to the doublet separations are $255.5$ and $32.46$\thinspace d, respectively, Tab.~\ref{tab:mod}). Side peaks with the same separation are present also at the only detected harmonic, $2f_0$, where, in addition $2f_0-2\fbj$ is present. The prewhitening sequence at $f_0$ and at $2f_0$ is illustrated in Fig.~\ref{fig.85370fsp} (left and right panels, respectively). One additional significant peak appears on the lower frequency side of $f_0$; its frequency is denoted with $f_{\rm u}$ in Fig.~\ref{fig.85370fsp}. It is significant (${\rm S/N}=10$, $13.1$\thinspace mmag), however no significant close peaks with $f_0-f_{\rm u}$ separation or significant combination frequencies involving $f_{\rm u}$ are detected in the spectrum (see also below). 

Side peaks with $\fbj$ separation also appear at combination frequencies with the first overtone, but not at $f_1$. Here, we only detect one side peak on higher frequency side, at $f_1+\fbd$, with amplitude of $15.0$\thinspace mmag ($\pbd=51.59$\thinspace d). We note that amplitude of the first overtone itself is lower than amplitude of the highest side peak at $f_0$. We note that $\fbd\approx 5\fbj$ (strictly $\fbd/\fbj=4.95$). Also, $f_{\rm u}\approx f_0-\fbd+\fbj$. In fact, $f_{\rm u}$ and $f_0-\fbd+\fbj$ are not resolved within the OGLE-IV data. Based on the data we have, we cannot determine whether $f_{\rm u}$ is an independent frequency corresponding to modulation, or it corresponds to the above combination of other modulation frequencies.

\begin{figure}
\centering
\resizebox{\hsize}{!}{\includegraphics{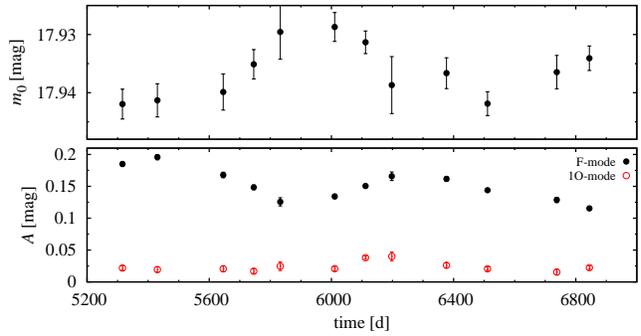}}
\caption{Mean stellar brightness (top) and mode amplitudes (bottom) for OGLE-BLG-RRLYR-22356 extracted with the time-dependent Fourier analysis ($\Delta t\approx 100$\thinspace d).}
\label{fig.85370feed}
\end{figure}

Fig.~\ref{fig.85370fsp} indicates that the fundamental mode is non-stationary. We have investigated the temporal variation of the fundamental and first overtone modes with the time-dependent Fourier analysis, first with large $\Delta t\approx 100$\thinspace d. In Fig.~\ref{fig.85370feed} we plot the mode amplitudes and mean stellar brightness. The fundamental mode is dominant and vary on a time-scale of $\sim 1000$\thinspace d. Its amplitude is strongly correlated with the mean stellar brightness. Lower the amplitude of the fundamental mode brighter the star. A long-term variation on a time-scale of $\sim 1000$\thinspace d is also visible for the first overtone mode.

We next investigated the suspected short-term modulations revealed in the frequency spectrum using lower $\Delta t\approx 10$\thinspace d. Results are rather noisy as data for this star are not too dense ($\rho\approx 310$) and the chosen $\Delta t$ is the lowest possible. Nevertheless, in the frequency spectra of the resulting 83-point time series and in the phase plots we confirm the modulation of the first overtone with \pbd\ and of the fundamental mode with \pbj. We cannot confirm modulation with $\pbt$ -- this requires a lower value of $\Delta t$, which is not possible with the data we have. The pulsation amplitude of the fundamental mode slowly decreases during the observations, which is also visible in Fig.~\ref{fig.85370feed}.

\subsection{OGLE-BLG-RRLYR-00951}

In this star pulsation in the fundamental mode dominates, however no sign of its modulation is detected (Tabs.~\ref{tab:dm} and \ref{tab:mod}). A clear, equally spaced triplet is visible at the first overtone frequency. Modulation period is $\pbj=142.7$\thinspace d and modulation amplitude is very low, of order of $5$\thinspace mmag. The triplet is also detected at $f_0+f_1$, while at $f_1-f_0$ a higher frequency component is missing. No other modulation components are detected. 

\begin{figure}
\centering
\resizebox{\hsize}{!}{\includegraphics{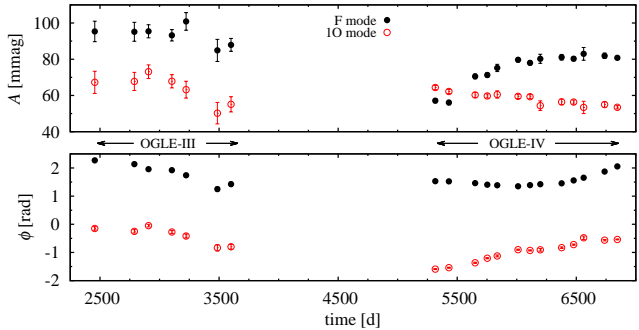}}
\caption{Time-dependent Fourier analysis for OGLE-BLG-RRLYR-00951 ($\Delta t\approx 100$\thinspace d).}
\label{fig.00951tdfd}
\end{figure}

We note that signals at $f_0$ and at $f_1$ are non-coherent and dominate the prewhitened frequency spectrum (${\rm S/N}=20.4$ and ${\rm S/N}=7.6$, respectively). It is a consequence of the long-term variation of fundamental and first overtone modes, clearly revealed in the time-dependent Fourier analysis -- Fig.~\ref{fig.00951tdfd}. Here we joined OGLE-III and OGLE-IV data, however we note that data sampling in OGLE-III is poor ($\rho=33$). Reasonable data are available only for four seasons. The gap between OGLE-III and OGLE-IV data is nearly 1600\thinspace days. No modulation is detected in OGLE-III data alone. The long-term variation of both modes is obvious and significant.

To get rid of non-coherent signal at $f_0$ and $f_1$ we conducted time-dependent prewhitening ($\Delta t\approx 140$\thinspace d) using OGLE-IV data only. Non-stationary signals at $f_0$ and $f_1$ were removed. We recovered all modulation components detected in the earlier analysis, but no new significant signal appeared. Because of low modulation amplitude time-dependent Fourier analysis cannot be used to study modulation with $\pbj$. With sufficiently low $\Delta t$, necessary for this analysis, errors of mode amplitudes are larger than the expected modulation amplitude.

\subsection{OGLE-BLG-RRLYR-11311}

Pulsation in the first overtone is dominant (Tab.~\ref{tab:dm}). Both at $f_0$ and at $f_1$ we detect doublets, with different separations however: one side peak is located at $f_1-\fbj$ ($32.4$\thinspace mmag, $\pbj=49.13$\thinspace d) and the other at $f_0+\fbd$ ($13.8$\thinspace mmag, $\pbd=87.1$\thinspace d). The side peaks with $\fbj$ separation are also detected at the combination frequencies, while side peak with $\fbd$ separation is detected only at $f_0$ and $2f_0$. Interestingly, we also find a significant (${\rm S/N}=5.9$) peak at $f_0+\fbd/2$ with amplitude of $6.7$\thinspace mmag. As in the case of OGLE-BLG-RRLYR-14915 (see comments in Section~\ref{sec:14915}) we interpret this signal as corresponding to sub-harmonic frequency.

\begin{figure}
\centering
\resizebox{\hsize}{!}{\includegraphics{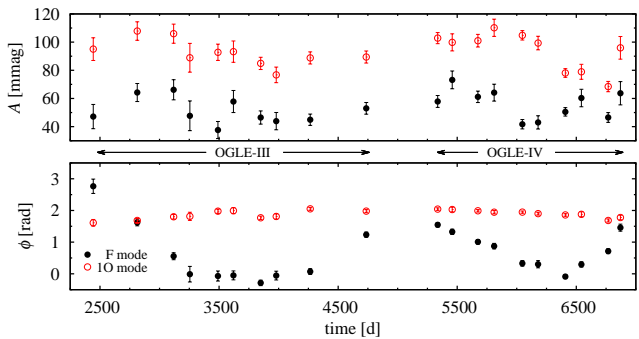}}
\caption{Time-dependent Fourier analysis for OGLE-BLG-RRLYR-11311 ($\Delta t\approx 150$\thinspace d).}
\label{fig.11311tdfd}
\end{figure}

The frequency spectrum of residual data is dominated by strong non-coherent signals at $f_0$ and at $f_1$ (${\rm S/N}=14.1$ and ${\rm S/N}=6.8$, respectively). Using the time-dependent Fourier analysis we confirm that both modes vary on a long time-scale -- Fig.~\ref{fig.11311tdfd}. It seems that long-term modulation of both modes, their amplitudes and phases, on a time-scale of $\sim 2000$\thinspace d occurs. More data are necessary to resolve the suspected modulation.

To get rid of the non-coherent signals at $f_0$ and at $f_1$ we conducted the time-dependent prewhitening ($\Delta t\approx 100$\thinspace d) using the OGLE-IV data only. Non-stationary signal at $f_1$ was removed and at $f_0$ was strongly reduced. We recovered all side peaks detected in the earlier analysis, but no new significant signal appeared. Because of low cadence of observations ($\rho=43/99$ in OGLE-III/IV) the data are insufficient to reveal radial mode modulations with $\pbj$ or $\pbd$ using time-dependent Fourier analysis. At the required $\Delta t$, errors of the determined amplitudes become larger than expected amplitude of modulation.

\subsection{OGLE-BLG-RRLYR-10728}

In this star fundamental mode dominates -- its mean amplitude is the highest in our sample. Two doublets with different separations are detected at $f_0$. The highest peak is located at $f_0+\fbj$ ($26.7$\thinspace mmag, $\pbj=98.7$\thinspace d). The highest peak on the lower frequency side of $f_0$ ($8.9$\thinspace mmag at $f_0-\fbd$) corresponds to modulation with period of $\pbd=313$\thinspace d. We cannot exclude that equidistant triplet with $\fbd$ separation is present -- unfortunately, $f_0+\fbd$ falls exactly at location of 1-year alias of unresolved signal at $f_0$. Also a triplet with $\fbj$ separation is possible, the peak at $f_0-\fbj$ is well visible, but weak (${\rm S/N}=3.4$).

Side peak with $\fbj$ separation is also detected at first overtone frequency (at $f_1-\fbj$, $8.6$\thinspace mmag). No side peaks with $\fbd$ separation are detected at $kf_1$. The modulation spectrum is rich for this star. In addition to side peaks with $\fbj$ and $\fbd$ separation also side peaks with $\fbj+\fbd$ separation appear at $2f_0$ and $3f_0$ (on the higher frequency side; also at $4f_0$ and $f_1-f_0$, but these peaks have ${\rm S/N}<4$) and also at $f_0$, $f_0+f_1$ and $2f_0+f_1$, but on the lower frequency side. Amplitudes of these peaks are all below $6$\thinspace mmag. In the prewhitened spectrum we find a weak residual power at $f_0$ (${\rm S/N}=5.5$) and at $f_1$ (${\rm S/N}=5$). The time-dependent Fourier analysis indicates that both amplitude and phase of the radial modes slowly vary within the OGLE-IV data. 

The OGLE-III data for this star are scarce: only for 4 weakly sampled seasons ($\rho=30$) reasonable data are available, separated form OGLE-IV observations by a large gap of nearly $1600$\thinspace days. Independent analysis of this data reveals that likely a sudden change in the pulsation properties of the star occurred in between OGLE-III and OGLE-IV observations. The pulsation periods as derived from OGLE-III data are $P_0=0.481690(4)$\thinspace d and $P_1=0.358121(3)$\thinspace d (cf. Tab.~\ref{tab:dm}). Hence, period of the first overtone decreased by nearly $0.003$\thinspace d, i.e., by $0.8$\thinspace per cent. Period of the fundamental mode decreased only slightly ($0.003$ per cent). As a result period ratio decreased from $0.7435$ (OGLE-III) to $0.7377$ OGLE-IV. The phase of fast change of first overtone period had to occur before OGLE-IV data, as within OGLE-IV the phase/amplitude changes are slow. In the OGLE-III data we detect only one side peak on the higher frequency side of $f_0$ with separation corresponding to modulation period of $92.3$\thinspace d (to be compared with $\pbj=98.7$\thinspace d detected in the OGLE-IV data). The nature of these intriguing changes will remain obscure as we lack the crucial observations in between OGLE-III and OGLE-IV.

\subsection{OGLE-BLG-RRLYR-04598}

This star has the highest period ratio in our sample. In addition $P_1/P_0$ is almost exactly $0.75$ indicating that $3\!:\!4$ resonance may play a role in this star. Pulsation in the first overtone dominates. At the location of the fundamental mode we detect two lower frequency side peaks, at $f_0-\fbj$ and at $f_0-\fbd$, with amplitudes $12.9$ and $10.8$\thinspace mmag, respectively ($\pbj=281$\thinspace d, $\pbd=85.6$\thinspace d). Side peaks at the same separations are also detected near combination frequency $f_0+f_1$, the one at $f_0+f_1-\fbd$ is very weak, however (${\rm S/N}=4.0$). No side peak is detected directly at first overtone frequency, we note however, that unresolved peak remains close to $f_1$ after prewhitening (${\rm S/N}=9.5$). Location of its one-year alias is unresolved with $f_1-\fbj$. There is also a weak detection at $f=2f_1-\fbj$ (${\rm S/N}=4$). Since there is no firm evidence of modulation of the first overtone the relevant part of Tab.~\ref{tab:mod} is not filled.

In OGLE-III data we also find two side peaks on lower frequency side of $f_0$, one of them at roughly the same separation ($\fbd$), the other farther away, the inverse of separation corresponds to $258$\thinspace d, to be compared with $\pbj=281$\thinspace d. This indicates that possible modulation with $\pbj$ is non-stationary on a longer time-scale.

\begin{figure}
\centering
\resizebox{\hsize}{!}{\includegraphics{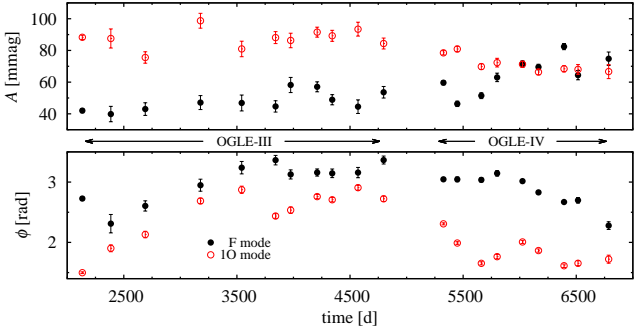}}
\caption{Time-dependent Fourier analysis for OGLE-BLG-RRLYR-04598 ($\Delta t\approx 130$\thinspace d).}
\label{fig.04598tdfd}
\end{figure}

In Fig.~\ref{fig.04598tdfd} we present a long-term variation of radial mode amplitudes and phases extracted with the time-dependent Fourier analysis ($\Delta t\approx 130$\thinspace d). Both modes vary on a long time-scale in an irregular fashion. Phase variation is more pronounced .The periods of both modes increase on average. A faster, $\sim$few hundred days irregular phase variation is also well visible. A slow trend of increasing amplitude of the fundamental mode and decreasing amplitude of the first overtone is apparent. 

Although the data are relatively dense ($\rho=146/375$ in OGLE-III/IV), the amplitude of suspected modulation is low. In addition, the irregular variability we see in Fig.~\ref{fig.04598tdfd} is of higher amplitude. Therefore, we cannot detect modulation using the time-dependent Fourier analysis with low $\Delta t$.

\subsection{OGLE-BLG-RRLYR-13442}

OGLE-BLG-RRLYR-13442 (marked with pentagon in Fig.~\ref{fig.pet}) is a special star. As reported by \cite{ogleiv_rrl_blg} the star switched pulsation mode between single-mode fundamental mode pulsation (RRab star, OGLE-III) to double-mode RRd pulsation (OGLE-IV). In the OGLE-IV data we detect three significant close peaks on the higher frequency side of $f_0$. The highest, located at $f_{\rm u1}\approx f_0+0.028$ has an amplitude of $29.4$\thinspace mmag (${\rm S/N}=13.4$), but no other significant peak with the same separation is detected. Only a weak (${\rm S/N}=3.3$) peak is present on higher frequency side of $f_0+f_1$. Similarly, at $f_{\rm u2}\approx f_0+0.032$ we find a much weaker peak (${\rm S/N}=5.1$), but again, no combination frequencies involving $f_{\rm u2}$ are detected in the spectrum. The other side peak, which we associate with modulation, is located at $f_0+\fbj$ ($14.0$\thinspace mmag, $\pbj=179.1$\thinspace d). A side peak with the same separation is detected also at $2f_0$. We find no significant side peak at the frequency of the `just born' first overtone. The frequency spectrum of residual data is dominated by strong unresolved signals at $f_0$ (${\rm S/N}=15.3$) and also at $f_1$ (${\rm S/N}=4.6$). 

Interestingly, side peak with $\fbj'\approx\fbj$ ($\pbj'=171.3$\thinspace d, equal to $\pbj$ within data resolution) separation and a peak at $f_{\rm u}'=f_{\rm u}$, are also detected in the OGLE-III data, in which we find no evidence for the presence of the first overtone. The side peak at $f_0+\fbj'$  has an amplitude of $29.0$\thinspace mmag. Despite a rather dramatic change of pulsation state -- the star switched pulsation mode from RRab to RRd -- the side peaks at the fundamental mode, their location and height, changed little.

\begin{figure}
\centering
\resizebox{\hsize}{!}{\includegraphics{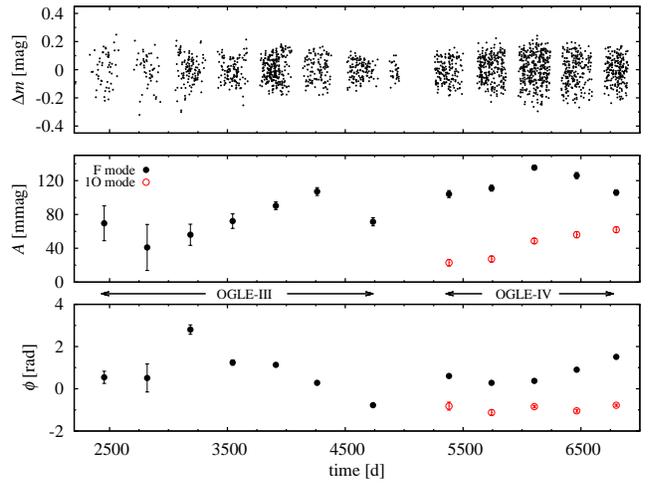}}
\caption{Data (top) and time-dependent Fourier analysis (middle and bottom) for OGLE-BLG-RRLYR-13442 (each point corresponds to one observing season).}
\label{fig.13442tdfd}
\end{figure}

Pulsation of this interesting star is clearly non-stationary. A lot of unresolved power remains in the frequency spectrum after prewhitening with significant frequencies. In Fig.~\ref{fig.13442tdfd} we plot a merged OGLE-III and OGLE-IV data (top) with season-to-season amplitudes and phases of the fundamental and first overtone modes extracted with the time-dependent Fourier analysis. The Figure suggests that pulsation may be modulated on a long-time scale of $\sim 1000$\thinspace days. Data are insufficient to reveal possible short-term modulation (with $\pbj/\pbd$) with the time-dependent Fourier analysis.

\subsection{OGLE-BLG-RRLYR-30986} 

For this star the data were collected only during the OGLE-IV phase. First overtone pulsation is dominant. Both at $f_0$ and at $f_1$, on their lower frequency side, we detect side peaks at equal separation, $\fbj$ (Tab.~\ref{tab:mod}). In addition at $f_1$ there is a peak at $f_1-2\fbj$. We interpret these side peaks as due to modulation of both modes with a common period of $\pbj=111.4$\thinspace d. Modulation of the fundamental mode is stronger ($51$\thinspace mmag at $f_0-\fbj$ to be compared with $14.4$\thinspace mmag at $f_1-\fbj$). Side peaks appear also at combination frequencies at the same separation. In the frequency spectrum of residual data a weak (${\rm S/N}=4.1$) signal at $2f_0$ remains. The data for this star are too sparse ($\rho\approx 98$) to apply the time-dependent Fourier analysis in order to investigate the modulation in more detail.

\subsection{OGLE-BLG-RRLYR-02530}

\begin{figure}
\centering
\resizebox{\hsize}{!}{\includegraphics{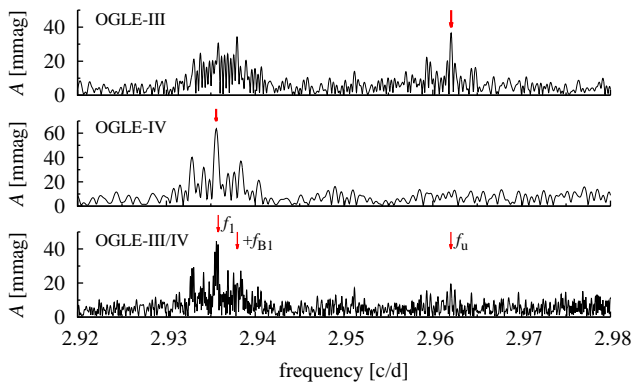}}
\caption{Fourier spectrum for OGLE-BLG-RRLYR-02530 in the frequency range around the first overtone frequency. Top panel for OGLE-III, middle panel for OGLE-IV and bottom panel for the merged data. Arrows mark the location of the first overtone frequency adopted in the respective catalogs. In the bottom panel we mark the first overtone frequency as adopted in this analysis and the location of modulation peak (visible in the OGLE-III data, top panel, and also in the OGLE-IV data, but only after prewhitening).}
\label{fig.02530fsp}
\end{figure}

We note a significant difference in the first overtone period given in the OGLE-III and in the OGLE-IV catalogs. In Fig.~\ref{fig.02530fsp} we explain the origin of the difference. There are three significant peaks in the OGLE-III data in the frequency range of interest. The highest, marked with arrow in the top panel ($f_{\rm u}$), was interpreted as corresponding to the first overtone. In OGLE-IV and in the merged OGLE-III and OGLE-IV data other, lower frequency peak is the highest. Following \cite{ogleiv_rrl_blg} we interpret this peak as corresponding to the first overtone. In fact, the peak interpreted as corresponding to the first overtone mode in the OGLE-III data is weak in OGLE-IV, and we do not find any combination frequencies involving $f_{\rm u}$. Therefore, we do not have a proof that this frequency originates form OGLE-BLG-RRLYR-02530. 

We detect a family of modulation peaks with the same separation ($\fbj$, $\pbj=469$\thinspace d) at both radial modes and their combination frequencies, on their higher frequency sides. The highest side peak is located at $f_0+\fbj$ and has an amplitude of $22.9$\thinspace mmag. The highest peak at the first overtone frequency ($f_1+\fbj$) has an amplitude of $18.7$\thinspace mmag. Interestingly at $f_0+f_1$ and $2f_0$ we also detect the quintuplet components, i.e. peaks with separations $+\fbj$ and $+2\fbj$. Data are too scarce to reveal the discussed modulation in the time-dependent Fourier analysis ($\rho=30/43$ in OGLE-III/IV)

\begin{figure}
\centering
\resizebox{\hsize}{!}{\includegraphics{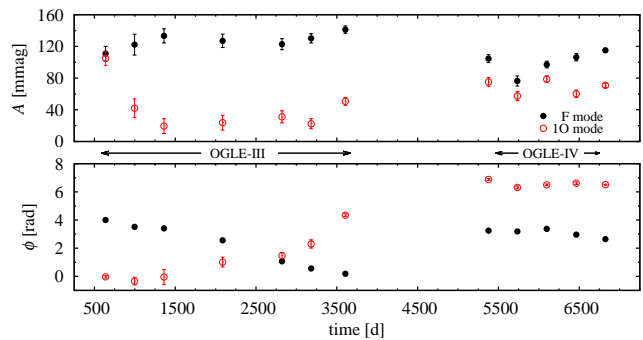}}
\caption{Time-dependent Fourier analysis for OGLE-BLG-RRLYR-02530 (each point corresponds to one observing season).}
\label{fig.02530tdfd}
\end{figure}

Fig.~\ref{fig.02530fsp} indicates that pulsation of the star is certainly non-stationary. In Fig.~\ref{fig.02530tdfd} we show the season-to-season variation of the radial modes. Both modes, their amplitudes and phases, strongly vary on a long time-scale. 

\subsection{OGLE-BLG-RRLYR-02862}

We find side peaks with the same separation, $\fbj$ ($\pbj=173.6$\thinspace d), at $f_0$ (equidistant triplet) and on the lower frequency side of $f_1$, $f_0+f_1$ and of $2f_1$ (weak). The triplet is asymmetric, the higher component at $f_0-\fbj$ has an amplitude of $28.6$\thinspace mmag, while the other component at $f_0+\fbj$ has an amplitude of only $7.9$\thinspace mmag (${\rm S/N}=4.3$). There are two additional, significant peaks in the frequency spectrum, one on the higher frequency side of $f_0$ (separation $\approx 0.038$\thinspace c/d), the other on its lower frequency side (separation $\approx 0.0072$\thinspace c/d), but we do not detect any other side peaks with these separations. Data for this star are insufficient to conduct more detailed analysis ($\rho\approx 34$ both in OGLE-III and in OGLE-IV). On a long time-scale variation of both radial modes is apparent and irregular. In particular during the OGLE-III the fundamental mode pulsation was dominant, while the first overtone dominates during the OGLE-IV observations.

\subsection{Other analysed stars}

In OGLE-BLG-RRLYR-02985, \mbox{-12660}, \mbox{-33181}, \mbox{-13804}, \mbox{-13302} and \mbox{-31786} (marked with circles in Fig.~\ref{fig.pet}) we do not find additional significant signals beyond $f_0$, $f_1$ and their combinations. We note that first four stars are located relatively close to the main progression formed by the majority of RRd stars in the Petersen diagram. Data for all these stars are rather sparse and presence of additional signal cannot be excluded. 

In frequency spectra of OGLE-BLG-RRLYR-05475, \mbox{-09965}, \mbox{-14031}, \mbox{-15012} and \mbox{-19451} (marked with squares in Fig.~\ref{fig.pet}), after prewhitening with radial mode frequencies and their combinations, we detect signal at first overtone frequency, unresolved with $f_1$ -- a signature of long-term variation of this mode. OGLE-BLG-RRLYR-05475 and OGLE-BLG-RRLYR-14031 are located within the main group of RRd stars (they were selected as primary targets).

OGLE-BLG-RRLYR-24137 (marked with triangle in Fig.~\ref{fig.pet}) turned out to be very unusual and interesting. It is likely a triple-mode RR~Lyrae star. Except fundamental and first overtone modes we detect a third mode, which is either the third radial overtone or a non-radial mode. Interestingly, the period doubling of the third mode is clearly detected. We discuss this star in detail in a separate publication \citep{smolec_tri}.

\section{Discussion}\label{sec:discussion}

\subsection{Modulation properties of the stars.}\label{ssec:dis_modp}

Modulation properties of the analysed stars are collected in Tab.~\ref{tab:mod}. In six stars the modulation is well established: either triplets are detected in the frequency spectrum or modulation is directly visible in the time-dependent Fourier analysis. In the latter case, doublets in the frequency spectrum are not an exception (e.g. in OGLE-BLG-RRLYR-05762, \mbox{-06283}, \mbox{-7393}, \mbox{-32462}). Therefore, we assume that doublets detected in 9 other stars are also due to modulation \citep[see also Section~\ref{sec:overview} and][]{alcock_rrab}. The other explanation, that additional close side peaks correspond to non-radial modes and their combinations with radial mode frequencies cannot be excluded, but in our opinion is less probable. 

In the adopted modulation picture, doublets are in fact triplets with highly asymmetric components. As pointed in Section~\ref{sec:overview} the detected triplets can indeed be highly asymmetric. To generate highly asymmetric triplets both amplitude and phase modulation of comparable strength is needed, as analysed by \cite{benko11}. This is indeed the case for the stars in which time-dependent Fourier analysis of modulation was possible (Figs.~\ref{fig.05762subsets}, \ref{fig.06283subsets}, \ref{fig.07393subsets} and \ref{fig.59913subsets}). The other factor necessary to generate the highly asymmetric triplet is a close to $\pm \pi/2$ phase-lag between the amplitude and phase modulations. If the offset is $0$ or $\pi$ a triplet should be symmetric. Unfortunately, the phased modulation curves we get for some stars are characterized by large scatter due to irregularities in the modulation and presence of other modulation periods. The lag between the amplitude and phase modulation cannot be determined accurately. Nevertheless we note that in the best case of OGLE-BLG-RRLYR-05762 and modulation of fundamental mode with $\pbj$ (Fig.~\ref{fig.05762subsets}) the lag is certainly very small. A triplet with comparable height of the modulation components ($53$ and $25$\thinspace mmag) is detected in its frequency spectrum. This is a rare case and, as analysed above, modulation we see in some stars and suspect in the others must be both amplitude and phase modulation, with a significant, close to $\pm \pi/2$, lag between the two modulations.

In 7 stars we detect a single modulation period, in 5 stars we detect two modulation periods and in 3 stars three modulation periods are detected. Altogether in 15 stars 26 modulation periods are found, of which 20 are connected with the fundamental mode, 14 are connected with the first overtone and 8 with both of the modes simultaneously. In six out of seven stars in which both modes are modulated, modulation of the fundamental mode is stronger (has higher amplitude). Modulation periods vary from few tenths to few hundred days, radial mode amplitudes may be modulated by a few to nearly hundred per cent. We do not find any significant correlations between e.g. the modulation periods and modulation amplitudes. We note however, that any statistics and description of modulation properties in our sample must be biased: the sample is small and not homogeneous. Data for different stars are of very different quality. Only in few densely sampled cases very detailed analysis of modulation was possible. Below we describe the most interesting phenomena found in the data.

{\bf Feedback effect between star brightness and pulsation mode amplitudes.} Due to radial mode modulation, the kinetic energy of pulsation varies in time. It is transferred to the thermal energy of the star and {\it vice versa}. The star adjusts its structure to the instantaneous mode amplitudes, which we observe through mean brightness variation. Since modulation we observe in analysed stars is typically dominant for one of the modes, the feedback effect we observe is the following: larger the amplitude of the dominant mode, smaller the mean stellar brightness. Then kinetic energy of pulsation drops, the mean brightness increases. The effect is best visible for OGLE-BLG-RRLYR-05762 ($A_0-m_0$, Fig.~\ref{fig.05762subsets}), OGLE-BLG-RRLYR-06283 ($A_0-m_0$, Fig.~\ref{fig.06283subsets}), OGLE-BLG-RRLYR-07393 ($A_1-m_0$, Fig.~\ref{fig.07393subsets}) and OGLE-BLG-RRLYR-22356 ($A_0-m_0$, Fig.~\ref{fig.85370feed}).

Properties of one radial mode are coupled with properties of the other mode. Both modes compete for the same driving mechanism and their amplitudes and phases are coupled through the cross-saturation effects. Therefore, modulation of one radial mode should be accompanied by modulation of the other mode with the same period. Some of the analysed stars seem to contradict the above statement, as modulation of only one mode is detected in the data (e.g. OGLE-BLG-RRLYR-09284, -04598) or the other mode is modulated, but with different period (e.g. OGLE-BLG-RRLYR-11311). We note however, that modulation amplitudes for each of the modes can differ significantly. In stars in which we find common modulation period for both modes, the modulation amplitudes can strongly differ (e.g. modulation with $\pbj$ in OGLE-BLG-RRLYR-05762, -07393 and in -02862, Tab.~\ref{tab:mod}) and rarely are of comparable strength. Because of the relatively high noise, inherent to ground-based observations, we will miss all modulations with amplitude below $\approx 3$\thinspace mmag (the exact limit depends on data quality for particular star). Moreover, modulation side peaks may be hidden in the bumps of excess power which we often find in the prewhitened spectra around radial mode frequencies. This excess power results from long-term irregular variations commonly present in the analysed stars (see below). 

In principle, when both modes are modulated simultaneously with the same period, the increase of one mode's amplitude should be accompanied by the decrease of other mode's amplitude. This is the case for the modulated double-overtone Cepheids analysed by \cite{mk09}. In these stars amplitudes of both modes are always modulated with a common (and long, $>700$\thinspace d) period and are anti-correlated \citep[see tab.~7 and figs.~15, 16 in ][]{mk09}. We do not find a clear evidence of such behaviour in our stars. When only one, common modulation period is detected, the data are insufficient to conduct the detailed analysis (our modulation periods are significantly shorter than in the case of the double-overtone Cepheids). When data are good enough, at least two modulation periods are detected and each of them is dominant for different radial mode: one for the fundamental, the other for the first overtone mode. One other factor that hinders the analysis of radial modes' amplitude correlations is non-stationary nature of the pulsation and of the modulation, which we discuss in more detail below.

The origin and mechanism of the detected modulations is unknown. We still lack the explanation for the single-periodic Blazhko effect in single-periodic RR~Lyrae stars, despite the great progress made recently thanks to {\it Kepler} and {\it CoRoT} observations \citep[see][for a review]{szabo14}. One of the crucial discoveries was detection of period doubling of the fundamental mode in the modulated RRab stars \citep{szabo10}, which motivated a new resonant model proposed by \cite{bk11} to explain the Blazhko effect. We do not find period doubling of any of the radial modes in our stars, but its detection may be difficult in the ground-based data. 

{\bf Long-term stability of the pulsations.} All of the investigated stars share a common feature: their pulsation properties are non-stationary. Amplitudes and phases of the radial modes vary irregularly on a long time-scales of few hundred, or thousands days. The long-term variations are clearly revealed in stars for which OGLE-III data are available. There are many examples in the paper, see e.g. Figs.~\ref{fig.05762tdfd}, \ref{fig.07393tdfd}, \ref{fig.09284tdfd}, \ref{fig.85370feed}, \ref{fig.00951tdfd}, \ref{fig.11311tdfd}, \ref{fig.04598tdfd} and \ref{fig.02530tdfd}. Both amplitude and periods of both radial modes vary on a long time-scale. This variation is most likely irregular, at least no obvious long-period modulation is clear in the discussed cases. These long-term irregular changes are reflected in the prewhitened frequency spectrum as close, but unresolved peaks at the location of radial modes. 

The short-term modulations reported in Tab.~\ref{tab:mod} are also not strictly periodic. The modulation side-peaks are non-coherent. Irregularities are also visible in the modulations revealed with the time-dependent Fourier analysis. In some cases the short-term modulation differs significantly in the OGLE-III and OGLE-IV data for the same star. Although OGLE-III data are less suitable for the study of short-term modulation, in some cases differences are obvious and significant. Example includes OGLE-BLG-RRLYR-07393 (see Tab.~\ref{tab:mod2} and Fig.~\ref{fig.07393tdfd}).  

{\bf Period changes.} Long-term phase variations are common to all analysed stars as pointed above. In the case of OGLE-BLG-RRLYR-05762, \mbox{-06283}, \mbox{-07393} and \mbox{-32462} we could analyse the radial mode's phase (period, eq.~\ref{eq:pc}) variation in more detail (Figs.~\ref{fig.05762subsets}, \ref{fig.06283subsets}, \ref{fig.07393subsets} and \ref{fig.59913subsets}, respectively). We note that phases of low amplitude of the radial mode are accompanied by fast phase change. In other words, mode of low amplitude is prone to fast period change. In particular, when amplitude of the mode drops to very small value a glitch in the phase variation is observed (OGLE-BLG-RRLYR-07393, Fig.~\ref{fig.07393subsets}, OGLE-BLG-RRLYR-32462, Fig.~\ref{fig.59913subsets}). The apparent phase discontinuity is an effect of our coarse sampling of the modulation properties (large $\Delta t$). We note that the correlation: lower the amplitude, faster the phase change is also apparent in RRc stars observed from space by {\it Kepler} space telescope. In case of these stars the phase variation can be traced continuously. The described effect is well visible in fig.~6 of \cite{pamsm14}. When amplitude of the mode drops, phase variation is rapid but continuous. 

In two cases we find an evidence for fast and linear change of first overtone period (OGLE-BLG-RRLYR-06283, Fig.~\ref{fig.06283subsets}, OGLE-BLG-RRLYR-05762, Fig.~\ref{fig.05762subsets}) on a time-scale shorter than expected as due to evolutionary changes. In other star, OGLE-BLG-RRLYR-10728, we find an evidence for an abrupt change of the first overtone period.

\subsection{Peculiarities of the analysed stars; the Petersen diagram}\label{ssec:dis_pet}

A glimpse at the Petersen diagram, presented in Fig.~\ref{fig.pet}, reveals that nearly all stars showing modulation are located off the main progression formed by the majority of RRd stars. At a given period of the fundamental mode their period ratios are either significantly smaller or significantly higher than in the majority of RRd stars. Two stars have typical period ratios, but are unusual for other reason -- they are likely members of a tidal stream crossing the Galactic bulge \citep{ogleiv_rrl_blg}.

We admit that we have not analysed all 173 RRd stars from the OGLE Galactic bulge collection. We only analysed those stars for which automatic prewhitening revealed close or unresolved signal at $f_0$ or at $f_1$ (see Sec.~\ref{sec:methods}). Later, then it was clear that all such stars have non-typical location in the Petersen diagram, we analysed all deviating stars and found additional two stars with modulation. We note that the OGLE-III data for all 92 RRd stars present in the OGLE-III catalog of variable stars were analysed while searching for non-radial modes \citep{netzel}. Some close side peaks or unresolved frequencies at $f_0$ and/or at $f_1$ were detected in a few stars. All of them are included in our sample and their period ratios are not typical. We cannot exclude that there are more stars which show modulation among the typical RRd stars, but we conclude that the fraction of stars showing modulation must be significantly lower there. Analysis of all 173 RRd stars using both OGLE-III and OGLE-IV data is ongoing. The present analysis indicates that at least $\sim 8.7$\thinspace per cent of RRd stars from the OGLE Galactic bulge collection are modulated, but their period ratios are not typical.

We note that location of these stars in the Petersen diagram presents a difficulty for stellar pulsation theory. Metallicity is the dominant parameter that affects the location of a star in the Petersen diagram. In particular the short period-low period ratio tail, well visible in Fig.~\ref{fig.pet} and absent in the LMC and SMC populations \citep{ogle_rr_lmc,ogle_rr_smc}, is explained by larger metallicity spread in the Galactic bulge population and higher metallicity of these stars -- see fig.~4 in \cite{ogle_rrl_blg} and fig.~2 in \cite{ogle_switch}. Explanation for stars off the main progression is missing. Their period ratios can also be matched with higher metallicity models \citep[see][]{smolec_tri}, but the reason why they deviate form the main progression in the Petersen diagram is unknown.

Not only period ratios are not typical in our sample. We note that in 7 stars ($\sim 47$\thinspace per cent) the fundamental mode dominates. If we consider only the stars located off the main progression in the Petersen diagram, the fraction is even higher ($\sim 54$\thinspace per cent). To the contrary, in the full sample of 173 Galactic bulge RRd stars the first overtone pulsation strongly dominates ($82.1$\thinspace per cent). Interestingly, one of the analysed stars, with the dominant fundamental mode pulsation, OGLE-BLG-RRLYR-13442, has switched the pulsation mode recently: from RRab to RRd pulsation. It is marked with a pentagon in Fig.~\ref{fig.pet} and its period ratio is also not typical. Its pulsation state is clearly non-stationary (like in other analysed stars). The same modulation side peaks were present both before the mode switch and after the mode switch. With the data we have, speculations about possible connection between non-typical period ratio, RRab$\rightarrow$RRd mode switch, and non-stationary modulations are premature. We think however, that the hypothesis that such connection exists is interesting and worth investigating. We plan a detailed analysis of long-term stability of radial mode properties in the full sample of RRd stars. Certainly, the stars that deviate from the main progression in the Petersen diagram require long-term monitoring. We note that there are two other Galactic bulge stars in OGLE database that switched the pulsation mode but in a reverse direction: from RRd to RRab \citep[OGLE-BLG-RRLYR-12245 and OGLE-BLG-RRLYR-07226, see][]{ogle_switch,ogleiv_rrl_blg}. Before the mode switch, their period ratios were typical.

\subsection{Modulation properties as compared to Blazhko effect reported in the literature}\label{ssec:dis_other}

The Blazhko effect, amplitude and phase modulation of radial modes, is commonly detected in single-periodic RR~Lyrae stars, both in RRab and in RRc stars. The effect is more frequent in RRab stars. Space photometry and analysis of top-quality ground-based data indicate that it may occur in up to $50$\thinspace per cent of stars \citep{benko10,benko14,jurcsik_kbs}. The frequency is likely lower in the case of RRc stars, up to $10$\thinspace per cent \citep[e.g.][]{mizerski,nagy}, but these stars are relatively little studied; in particular we lack space photometry for modulated RRc stars.  We note that in the ground-based data, multiple modulation periods are rare. Twenty such stars are listed in the \cite{skarka} catalog of the Galactic field Blazhko stars \citep[two modulation periods are commonly detected in the space photometry, see][]{benko14}. The modulation observed in single-periodic stars is not strictly periodic. The consecutive modulation cycles can differ strongly, which is best visible in the space photometry for these stars  \citep[see e.g.][]{benko14,gugg}. In some stars the modulation is not stable on the long time-scale. The best studied example is RR~Lyr, in which modulation nearly vanished recently. Whether it will reappear in a rhythm of a postulated 4 year cycle is not clear at the moment \citep{leborgne}. We lack a thorough and detailed analysis of long-term stability of the Blazhko modulation. The coupled OGLE-III and OGLE-IV data seem best for this purpose.

Our results also indicate that the fundamental mode is more prone to modulation (Tab.~\ref{tab:mod} and discussion above). In our stars multi-periodic modulation is very frequent. Only in seven stars we detect a single modulation period. Modulation we detect in our stars is clearly irregular and non-stationary.

We also note that Blazhko effect was previously detected in double-mode 1O+2O Cepheids -- see \cite{mk09} for detailed analysis. These stars have very different properties however, as pointed out above.

Just as we finished our analysis, the paper by \cite{jurcsik_bl} was published, in which they announce the discovery of the Blazhko effect in four RRd stars in M3 cluster. Their results are vey similar to ours. Their stars have non-typical period ratios. In three of them the two radial modes are modulated with different periods \citep[see fig. 3 and tab. 1 in][]{jurcsik_bl}.

\section{Summary}\label{sec.summary}

We have analysed the OGLE data for RRd stars of the Galactic bulge collection suspected of Blazhko-type modulation and selected by visual inspection of the data or by automatic analysis of the frequency spectrum. In the case of 13 stars we found a modulation of at least one radial mode (triplets or close doublets in the frequency spectrum). Since eleven of these stars are located off the main progression of RRd stars in the Petersen diagram, we decided to check all of the remaining deviating stars. Additional two stars with modulation of the radial modes were found and thus the full sample counts 15 stars. For all these stars we analysed long-term stability of pulsation and modulation properties using OGLE-IV and OGLE-III data, if available. Most important findings and conclusions of this study are:
\begin{itemize}
\item RRd stars located off the main progression in the Petersen diagram are prone to modulation. Thirteen out of twenty three deviating stars show the modulation.
\item The short-term modulation is typically multi-periodic, with modulation periods between few tenths to few hundred days; Modulation is not strictly periodic.
\item In most of the stars we detect a long-term, irregular variation of radial mode amplitudes and periods on top of the short-term modulation.
\item In several stars we find a feedback between the amplitude of the dominant radial mode and mean stellar brightness: lower the pulsation amplitude, brighter the star.
\end{itemize}
Two stars deserve a special mention:
\begin{itemize}
\item In OGLE-BLG-RRLYR-07393 we find additional non-radial mode excited, with characteristic period ratio to first overtone, around $\sim 0.61$. It is a member of new class of radial--non-radial RR~Lyrae stars revealed recently \citep[][and references therein]{pamsm14}. We find an evidence that the non-radial mode is modulated with the same period as the radial modes.
\item OGLE-BLG-RRLYR-13442 experienced a pulsation mode switch, from RRab to RRd pulsation, in between the third and fourth phases of the OGLE project. Its suspected modulation, manifesting through close doublets in the frequency spectrum, was not affected significantly by the mode switching event.
\end{itemize}

What is the mechanism behind modulation in RRd stars? Is it the same mechanism as in the case of the Blazhko effect in single-periodic stars? What is the mechanism behind the multi-periodic modulation? These are all open questions. The data presented in this paper will help to verify the theories to come. As: ({\it i}) the majority of the discussed stars are located off the main progression of RRd stars in the Petersen diagram, ({\it ii}) one of them experienced an RRab$\rightarrow$RRd mode switch, ({\it iii}) in a large fraction of these stars pulsation in the fundamental mode dominates (in contrast to the stars with the typical period ratios) and ({\it iv}) pulsation properties of these stars are non-stationary we form a loose hypothesis: The investigated stars are in a transient pulsation state that possibly follows the RRab$\rightarrow$ RRd mode switching. To test the hypothesis detailed analysis of all RRd stars is necessary and ongoing. More importantly, long-term monitoring of the discussed stars is needed to shed more light on their intriguing nature.

\section*{Acknowledgments}

We are grateful to Pawe\l{} Moskalik for stimulating discussions, reading the manuscript and detailed comments. Fruitful conversation with Wojtek Dziembowski is also acknowledged.

This research is supported by the Polish National Science Centre through grant DEC-2012/05/B/ST9/03932 and by the Polish Ministry of Science and Higher Education through the program ``Ideas Plus'' award No. IdP2012 000162. The OGLE project has received funding from the European Research Council under the European Community's Seventh Framework Programme (FP7/2007-2013)/ERC grant agreement no. 246678 to AU.

\appendix
\section[]{On-line only Tables with light curve solutions}
In the Appendix we present the full light curve solutions of OGLE-IV data for all analysed stars. Tables, available in the on-line version of the Journal, contain frequency identification, frequency value, amplitude and phase, all with standard errors. For a sample see Tab.~\ref{tab:appsample}. 

\bsp

\clearpage

\begin{table*}
\centering
\caption{Light curve solution for OGLE-BLG-RRLYR-00951. The consecutive columns contain frequency id., frequency value, amplitude with standard error, phase with standard error and remarks. Entries are sorted by increasing frequency. In case of frequency values the error of the last two digits of independent frequencies is given in parenthesis. Asterisk at amplitude value indicates that corresponding peak is non-stationary. `bl' in the `remarks' column indicates a frequency of the Blazhko modulation in case no significant peak was detected in the frequency spectrum directly at $f=f_{\rm B}$. Amplitude and phase are not given then. For some frequencies a signal-to-noise ratio is given in the remarks column, if it is lower than 4.}
\label{tab:freqs00951}
\begin{tabular}{lr@{.}lrrrrr}
freq. id & \multicolumn{2}{c}{$f$\thinspace [d$^{-1}$]}& $A$\thinspace [mmag] & $\sigma$ & $\phi$ [rad] & $\sigma$ & remarks\\ 
\hline  
 $\fbj$         &  0&007010(15)  &      &      &        &      & bl\\  
 $f_1-f_0-\fbj$ &  0&6718715     &  4.3 & 0.4  &  2.57  & 0.59 &\\
 $f_1-f_0$      &  0&6788813     &  6.0 & 0.4  &  2.18  & 0.13 &\\
 $f_0$          &  1&8360933(18) &*73.4 & 0.4  &  5.64  & 0.07 &\\  
 $f_1-\fbj$     &  2&5079647     &  5.5 & 0.4  &  3.23  & 0.59 &\\
 $f_1$          &  2&5149745(21) &*58.0 & 0.4  &  1.11  & 0.08 &\\
 $f_1+\fbj$     &  2&5219844     &  4.5 & 0.4  &  1.89  & 0.59 &\\
 $2f_0$         &  3&6721865     &  7.2 & 0.4  &  0.89  & 0.16 &\\
 $f_0+f_1-\fbj$ &  4&3440580     &  1.7 & 0.4  &  5.03  & 0.64 & 3.9\\ 
 $f_0+f_1$      &  4&3510678     & 14.0 & 0.4  &  3.13  & 0.11 &\\
 $f_0+f_1+\fbj$ &  4&3580776     &  3.3 & 0.4  &  3.35  & 0.60 &\\
 $2f_1$         &  5&0299491     & 11.1 & 0.4  &  5.22  & 0.17 &\\
 $f_0+2f_1$     &  6&8660423     &  2.3 & 0.4  &  0.87  & 0.26 &\\
 $3f_1$         &  7&5449236     &  2.9 & 0.4  &  2.84  & 0.29 &\\
\hline
\end{tabular}
\end{table*}

\begin{table*}
\centering
\caption{Same as Tab.~\ref{tab:freqs00951} for OGLE-BLG-RRLYR-02530.}
\label{tab:freqs02530}
\begin{tabular}{lr@{.}lrrrrr}
freq. id & \multicolumn{2}{c}{$f$\thinspace [d$^{-1}$]}& $A$\thinspace [mmag] & $\sigma$ & $\phi$ [rad] & $\sigma$ & remarks\\ 
\hline  
         $\fbj$ &  0&002133(14)  &      &      &      &      & bl \\
      $f_1-f_0$ &  0&7783911     &   8.2 & 1.5 &  2.07 & 0.33 &\\
          $f_0$ &  2&1572025(44) &  95.6 & 1.6 &  6.09 & 0.17 &\\
     $f_0+\fbj$ &  2&1593352     &  22.9 & 1.6 &  2.12 & 0.48 &\\
          $f_1$ &  2&9355936(56) & *67.3 & 1.6 &  1.65 & 0.22 &\\
     $f_1+\fbj$ &  2&9377263     &  18.7 & 1.7 &  5.78 & 0.50 &\\
     $f_{\rm u}$ &  2&962149(35)  & *12.5 & 1.5 &  0.02 & 1.36  &\\
   $2f_0+2\fbj$ &  4&3186705     &   8.5 & 1.5 &  1.62 & 0.98 &\\
      $f_0+f_1$ &  5&0927961     &  18.4 & 1.6 &  3.87 & 0.29 &\\
 $f_0+f_1+\fbj$ &  5&0949288     &  13.0 & 1.6 &  1.33 & 0.45 &\\
$f_0+f_1+2\fbj$ &  5&0970616     &  10.5 & 1.6 &  0.47 & 0.95 &\\
    $2f_1+\fbj$ &  5&8733199     &   8.4 & 1.5 &  3.51 & 0.58 &\\
$f_0+2f_1+\fbj$ &  8&0305224     &  10.8 & 1.5 &  5.86 & 0.54 &\\
\hline
\end{tabular}
\end{table*}

\begin{table*}
\centering
\caption{Same as Tab.~\ref{tab:freqs00951} for OGLE-BLG-RRLYR-02862.}
\label{tab:freqs02862}
\begin{tabular}{lr@{.}lrrrrr}
freq. id & \multicolumn{2}{c}{$f$\thinspace [d$^{-1}$]}& $A$\thinspace [mmag] & $\sigma$& $\phi$ [rad] & $\sigma$& remarks\\ 
\hline  
  $\fbj$        &  0&005762(16) &       &     &       &      & bl\\ 
  $f_1-f_0$      &  0&8067095   &  11.5 & 1.5 &  2.12 & 0.36 &\\
  $f_{\rm u1}$    &  2&167575(41) &  11.6 & 1.6 &  3.53 & 1.61 &\\
  $f_0-\fbj$    &  2&1690406   &  28.6 & 1.5 &  5.82 & 0.62 &\\
  $f_0$         &  2&1748023(65)& *50.7 & 1.5 &  5.52 & 0.25 &\\
  $f_0+\fbj$    &  2&1805639   &   7.9 & 1.5 &  2.34 & 0.71 &\\
  $f_{\rm u2}$    &  2&212996(51) &   9.7 & 1.5 &  3.03 & 1.97 &\\
  $f_1-\fbj$    &  2&9757501   &  10.3 & 1.5 &  0.10 & 0.66 &\\
  $f_1$         &  2&9815118(56)&  81.4 & 1.5 &  2.01 & 0.22 &\\
  $2f_0$        &  4&3496046   &  24.4 & 1.5 &  2.73 & 0.50 &\\
  $f_1+f_0-\fbj$ & 5&1505524   &   9.8 & 1.5 &  3.50 & 0.68 &\\
  $f_1+f_0$      & 5&1563140   &  10.6 & 1.5 &  3.28 & 0.38 &\\
  $2f_1-\fbj$   &  5&9572619   &   6.6 & 1.5 &  4.77 & 0.77 &\\
  $2f_1$        &  5&9630235   &  12.8 & 1.5 &  0.61 & 0.45 &\\
  $2f_0+f_1$    &  7&3311163   &   6.0 & 1.5 &  1.77 & 0.61 & 3.5\\
\hline
\end{tabular}
\end{table*}

\begin{table*}
\centering
\caption{Same as Tab.~\ref{tab:freqs00951} for OGLE-BLG-RRLYR-04598.}
\label{tab:freqs04598}
\begin{tabular}{lr@{.}lrrrrr}
freq. id & \multicolumn{2}{c}{$f$\thinspace [d$^{-1}$]}& $A$\thinspace [mmag] & $\sigma$ & $\phi$ [rad] & $\sigma$ & remarks\\ 
\hline  
  $\fbj$        &  0&003555(18)  &      &     &       &      & bl\\  
  $\fbd$        &  0&011685(21)  &      &     &       &      & bl\\
  $f_1-f_0$     &  0&7209096     & 10.9 & 0.7 &  5.42 & 0.18 &\\
  $f_0-\fbj$    &  2&1596111     & 12.9 & 0.7 &  3.94 & 0.66 &\\
  $f_0-\fbd$    &  2&1514809     & 10.8 & 0.7 &  5.08 & 0.78 &\\
  $f_0$         &  2&1631660(29) & 64.9 & 0.7 &  5.51 & 0.11 &\\
  $f_1$         &  2&8840756(30) &*71.0 & 0.7 &  4.46 & 0.11 &\\
  $2f_0$        &  4&3263319     &*18.2 & 0.7 &  0.29 & 0.22 &\\
  $f_1+f_0-\fbd$ &  5&0355564     &  3.7 & 0.7 &  5.51 & 0.80 &\\
  $f_1+f_0-\fbj$ &  5&0436866     &  5.2 & 0.7 &  5.15 & 0.67 &\\
  $f_1+f_0$      &  5&0472415     & 19.7 & 0.7 &  5.99 & 0.15 &\\
  $2f_1-\fbj$    &  5&7645962     &  3.2 & 0.7 &  6.05 & 0.71 &\\
  $2f_1$         &  5&7681511     &  9.4 & 0.7 &  0.39 & 0.23 &\\
  $3f_0$         &  6&4894979     &  3.2 & 0.7 &  1.58 & 0.38 &\\
  $f_1+2f_0$     &  7&2104075     &  9.2 & 0.7 &  0.54 & 0.25 &\\
  $2f_1+f_0$     &  7&9313171     &  6.7 & 0.7 &  1.05 & 0.26 &\\
  $3f_1$         &  8&6522267     &  4.6 & 0.7 &  1.95 & 0.36 &\\
  $f_1+3f_0$     &  9&3735735     &  3.4 & 0.7 &  2.35 & 0.39 &\\
  $2f_1+2f_0$    & 10&0944831     &  3.5 & 0.7 &  2.68 & 0.35 &\\
\hline
\end{tabular}
\end{table*}

\begin{table*}
\centering
\caption{Same as Tab.~\ref{tab:freqs00951} for OGLE-BLG-RRLYR-05762.}
\label{tab:freqs05762}
\begin{tabular}{lr@{.}lrrrrr}
freq. id & \multicolumn{2}{c}{$f$\thinspace [d$^{-1}$]}& $A$\thinspace [mmag] & $\sigma$ & $\phi$ [rad] & $\sigma$ & remarks\\ 
\hline  
         $\fbj$ &  0&0102097(16) &  6.7  & 0.3  &   4.93 & 0.07 &\\    
         $\fbt$ &  0&012327(12)  &       &      &        &      & bl\\ 
         $\fbd$ &  0&0434793(73) &  2.2  & 0.3  &   4.44 & 0.31 &\\    
 $f_1-f_0-\fbt$ &  0&72259893    &  2.3  & 0.3  &   2.70 & 0.47 &\\
 $f_1-f_0-\fbj$ &  0&72471593    &  2.0  & 0.3  &   5.32 & 0.18 &\\
$f_1-f_0-\fbt+\fbj$ &  0&73280865    &  1.8  & 0.3  &   4.64 & 0.48 &\\
      $f_1-f_0$ &  0&73492565    & *7.5  & 0.3  &   0.75 & 0.08 &\\
 $f_1-f_0+\fbj$ &  0&74513537    &  3.6  & 0.3  &   3.69 & 0.12 &\\
     $f_0-\fbj$ &  2&13434878    &*53.0  & 0.3  &   3.87 & 0.06 &\\
          $f_0$ &  2&14455850(58)& 94.5  & 0.3  &   0.46 & 0.02 &\\
     $f_0+\fbj$ &  2&15476822    & 25.0  & 0.3  &   4.94 & 0.07 &\\
$f_0+\fbd-\fbj$ &  2&17782804    &  3.2  & 0.3  &   4.54 & 0.30 &\\
     $f_0+\fbd$ &  2&18803776    &  7.1  & 0.3  &   3.82 & 0.28 &\\
     $f_1-\fbt$ &  2&86715743    &  7.3  & 0.3  &   1.58 & 0.45 &\\
     $f_1-\fbj$ &  2&86927442    &  5.6  & 0.3  &   5.94 & 0.10 &\\
          $f_1$ &  2&8794841(17) &*51.5  & 0.3  &   0.54 & 0.07 &\\
   $2f_0-2\fbj$ &  4&26869756    &  4.0  & 0.3  &   5.21 & 0.14 &\\
    $2f_0-\fbj$ &  4&27890728    &  5.4  & 0.3  &   0.21 & 0.09 &\\
         $2f_0$ &  4&28911700    &*44.3  & 0.3  &   2.67 & 0.05 &\\
    $2f_0+\fbj$ &  4&29932672    &  4.1  & 0.3  &   1.85 & 0.11 &\\
    $2f_0+\fbd$ &  4&33259625    &  8.3  & 0.3  &   5.93 & 0.28 &\\
 $f_0+f_1-\fbt$ &  5&01171593    &  2.2  & 0.3  &   0.30 & 0.47 &\\
 $f_0+f_1-\fbj$ &  5&01383292    &*11.6  & 0.3  &   0.66 & 0.09 &\\
      $f_0+f_1$ &  5&02404264    &  9.4  & 0.3  &   3.54 & 0.08 &\\
 $f_0+f_1+\fbj$ &  5&03425236    &  3.2  & 0.3  &   1.79 & 0.13 &\\
    $2f_1-\fbt$ &  5&74664157    &  1.9  & 0.3  &   4.72 & 0.48 &\\
         $2f_1$ &  5&75896829    &  3.8  & 0.3  &   3.82 & 0.15 &\\
    $3f_0-\fbj$ &  6&42346577    & *8.7  & 0.3  &   2.43 & 0.09 &\\
         $3f_0$ &  6&43367549    &*16.8  & 0.3  &   5.70 & 0.07 &\\
    $3f_0+\fbj$ &  6&44388522    &  4.2  & 0.3  &   3.90 & 0.12 &\\
    $3f_0+\fbd$ &  6&47715475    &  4.5  & 0.3  &   2.33 & 0.29 &\\
$2f_0+f_1-2\fbj$ &  7&14818170    &  2.4  & 0.3  &   1.14 & 0.18 &\\
 $2f_0+f_1-\fbj$ &  7&15839142    &  4.4  & 0.3  &   4.42 & 0.11 &\\
      $2f_0+f_1$ &  7&16860114    & 10.8  & 0.3  &   0.01 & 0.08 &\\
 $2f_0+f_1+\fbd$ &  7&21208040    &  1.3  & 0.3  &   2.79 & 0.36 &\\
 $f_0+2f_1-\fbj$ &  7&89331707    &  2.9  & 0.3  &   3.82 & 0.17 &\\
     $4f_0-\fbj$ &  8&56802427    &  2.9  & 0.3  &   5.68 & 0.14 &\\
          $4f_0$ &  8&57823399    &*11.7  & 0.3  &   2.13 & 0.09 &\\
     $4f_0+\fbj$ &  8&58844371    &  1.5  & 0.3  &   1.25 & 0.23 &\\
     $4f_0+\fbd$ &  8&62171325    &  4.0  & 0.3  &   4.67 & 0.29 &\\
 $3f_0+f_1-\fbt$ &  9&30083292    &  1.7  & 0.3  &   0.42 & 0.49 &\\
 $3f_0+f_1-\fbj$ &  9&30294992    & *5.4  & 0.3  &   0.11 & 0.11 &\\
      $3f_0+f_1$ &  9&31315964    &  3.3  & 0.3  &   2.77 & 0.13 &\\
     $5f_0-\fbj$ & 10&71258277    &  1.9  & 0.3  &   2.12 & 0.20 &\\
          $5f_0$ & 10&72279249    & *5.9  & 0.3  &   5.12 & 0.12 &\\
     $5f_0+\fbj$ & 10&73300221    &  1.5  & 0.3  &   3.73 & 0.24 &\\
     $5f_0+\fbd$ & 10&76627175    &  1.7  & 0.3  &   1.53 & 0.33 &\\
$4f_0+f_1-2\fbj$ & 11&43729870    &  2.0  & 0.3  &   6.01 & 0.20 &\\
 $4f_0+f_1-\fbj$ & 11&44750842    &  2.7  & 0.3  &   3.35 & 0.16 &\\
      $4f_0+f_1$ & 11&45771814    &  2.7  & 0.3  &   5.42 & 0.15 &\\
$3f_0+2f_1-\fbj$ & 12&18243406    &  1.4  & 0.3  &   3.42 & 0.25 &\\
          $6f_0$ & 12&86735099    & *3.3  & 0.3  &   1.66 & 0.16 &\\
     $6f_0+\fbd$ & 12&91083025    &  1.8  & 0.3  &   3.94 & 0.34 &\\
     $7f_0+\fbd$ & 15&05538875    &  1.3  & 0.3  &   0.98 & 0.38 &\\
 $5f_0+f_1-\fbj$ & 13&59206692    & *2.4  & 0.3  &   6.02 & 0.18 &\\
      $5f_0+f_1$ & 13&60227664    &  1.6  & 0.3  &   1.99 & 0.22 &\\
          $7f_0$ & 15&01190949    &  1.9  & 0.3  &   4.47 & 0.22 &\\
 $6f_0+f_1-\fbj$ & 15&73662541    &  1.1  & 0.3  &   3.00 & 0.30 &\\
          $8f_0$ & 17&15646799    &  1.3  & 0.3  &   1.27 & 0.29 &\\
 $7f_0+f_1-\fbj$ & 17&88118391    &  1.3  & 0.3  &   5.91 & 0.28 &\\
\hline
\end{tabular}
\end{table*}

\begin{table*}
\centering
\caption{Same as Tab.~\ref{tab:freqs00951} for OGLE-BLG-RRLYR-06283.}
\label{tab:freqs06283}
\begin{tabular}{lr@{.}lrrrrr}
freq. id & \multicolumn{2}{c}{$f$\thinspace [d$^{-1}$]}& $A$\thinspace [mmag] & $\sigma$ & $\phi$ [rad] & $\sigma$ & remarks\\ 
\hline  
       $\fbj$ &  0&0030046(29) &  4.9  & 0.4 &    2.72 & 0.12 &\\
     $f_1-f_0$ &  0&6977648     & *8.3  & 0.3 &    1.95 & 0.08 &\\
$f_1-f_0+\fbj$ &  0&7007694     &  7.4  & 0.3 &    2.85 & 0.12 &\\
    $2f_0-f_1$ &  1&2818661     &  3.3  & 0.3 &    1.45 & 0.15 &\\
    $f_0-\fbj$ &  1&9766263     &*37.0  & 0.3 &    2.45 & 0.10 &\\
         $f_0$ &  1&9796309(15) & 52.4  & 0.3 &    3.36 & 0.06 &\\
         $f_1$ &  2&6773956(12) &*62.5  & 0.3 &    5.84 & 0.05 &\\
    $2f_1-f_0$ &  3&3751604     & *3.8  & 0.3 &    5.63 & 0.13 &\\
   $2f_0-\fbj$ &  3&9562572     &  5.0  & 0.3 &    2.32 & 0.13 &\\
        $2f_0$ &  3&9592618     & 13.8  & 0.3 &    2.41 & 0.12 &\\
$f_1+f_0-\fbj$ &  4&6540219     & 11.2  & 0.3 &    4.38 & 0.11 &\\
     $f_1+f_0$ &  4&6570265     &*17.8  & 0.3 &    5.34 & 0.07 &\\
        $2f_1$ &  5&3547913     &*11.0  & 0.3 &    3.10 & 0.10 &\\
   $3f_0-\fbj$ &  5&9358881     &  2.1  & 0.3 &    1.15 & 0.22 &\\
        $3f_0$ &  5&9388926     &  3.7  & 0.3 &    2.18 & 0.19 &\\
    $f_1+2f_0$ &  6&6366574     &*10.9  & 0.3 &    4.45 & 0.12 &\\
$2f_1+f_0-\fbj$ &  7&3314176     &  2.9  & 0.3 &    0.68 & 0.17 &\\
     $2f_1+f_0$ &  7&3344222     &  4.6  & 0.3 &    1.67 & 0.12 &\\
        $3f_1$ &  8&0321869     & *3.0  & 0.3 &    5.77 & 0.16 &\\
$f_1+3f_0-\fbj$ &  8&6132837     &  1.2  & 0.3 &    3.55 & 0.30 &\\
    $f_1+3f_0$ &  8&6162883     &  2.5  & 0.3 &    4.70 & 0.21 &\\
   $2f_1+2f_0$ &  9&3140530     &  2.5  & 0.3 &    0.83 & 0.17 &\\
    $3f_1+f_0$ & 10&0118178     &  1.1  & 0.3 &    5.04 & 0.28 &\\
   $3f_1+2f_0$ & 11&9914487     &  1.2  & 0.3 &    4.24 & 0.29 &\\
\hline
\end{tabular}
\end{table*}

\begin{table*}
\centering
\caption{Same as Tab.~\ref{tab:freqs00951} for OGLE-BLG-RRLYR-07393.}
\label{tab:freqs07393}
\begin{tabular}{lr@{.}lrrrrr}
freq. id & \multicolumn{2}{c}{$f$\thinspace [d$^{-1}$]}& $A$\thinspace [mmag] & $\sigma$ & $\phi$ [rad] & $\sigma$ & remarks\\ 
\hline
            $\fbd$ &  0&0030971(43) &       &      &         &      &bl\\
            $\fbj$ &  0&0047654(39) &  8.0  & 0.4  &   5.21  & 0.16 &\\
           $2\fbd$ &  0&0061942     &  2.7  & 0.4  &   2.80  & 0.37 &\\
$f_1-f_0-\fbj-\fbd$ &  0&7302881     & *3.2  & 0.5  &   3.14  & 0.23 &\\
     $f_1-f_0-\fbj$ &  0&7333852     & 13.1  & 0.5  &   5.97  & 0.16 &\\
     $f_1-f_0-\fbd$ &  0&7350536     & *4.6  & 0.5  &   5.24  & 0.21 &\\
          $f_1-f_0$ &  0&7381507     &  3.3  & 0.5  &   4.41  & 0.21 &\\
              $f_0$ &  2&1612715(30) & 44.7  & 0.5  &   0.99  & 0.11 &\\
         $f_0+\fbd$ &  2&1643686     &*34.4  & 0.5  &   5.30  & 0.15 &\\
         $f_0+\fbj$ &  2&1660369     &  3.8  & 0.4  &   6.26  & 0.22 &\\
         $f_{\rm u}$ &  2&891057(34)  &  4.6  & 0.4  &   3.26  & 1.31 &\\
        $f_1-2\fbd$ &  2&8932279     &  5.5  & 0.5  &   5.10  & 0.36 &\\
         $f_1-\fbj$ &  2&8946567     & 50.8  & 0.5  &   1.67  & 0.12 &\\
         $f_1-\fbd$ &  2&8963250     & 11.0  & 0.5  &   3.25  & 0.20 &\\
    $f_1-\fbj+\fbd$ &  2&8977538     & 13.5  & 0.5  &   5.88  & 0.20 &\\
              $f_1$ &  2&8994221(25) &*56.9  & 0.5  &   5.17  & 0.10 &\\
    $f_1+\fbj-\fbd$ &  2&9010905     &  4.3  & 0.5  &   5.85  & 0.30 &\\
         $f_1+\fbd$ &  2&9025192     & 21.5  & 0.5  &   5.81  & 0.18 &\\
         $f_1+\fbj$ &  2&9041876     &  4.7  & 0.5  &   3.67  & 0.25 &\\
    $2f_1-f_0-\fbj$ &  3&6328074     &  2.8  & 0.4  &   0.42  & 0.24 &\\
             $2f_0$ &  4&3225429     &  3.1  & 0.5  &   4.73  & 0.27 &\\
        $2f_0+\fbd$ &  4&3256400     &  4.3  & 0.5  &   2.74  & 0.23 &\\
       $2f_0+2\fbd$ &  4&3287371     &  3.4  & 0.5  &   0.15  & 0.32 &\\
    $f_{\rm x}-\fbd$ &  4&7018026     &  2.5  & 0.5  &   3.27  & 1.67 &\\
         $f_{\rm x}$ &  4&704900(42)  &  2.9  & 0.5  &   2.14  & 1.65 &\\
     $f_1+f_0-\fbj$ &  5&0559282     &  6.5  & 0.5  &   5.30  & 0.18 &\\
     $f_1+f_0-\fbd$ &  5&0575965     &  2.6  & 0.5  &   0.48  & 0.32 &\\
$f_1+f_0-\fbj+\fbd$ &  5&0590253     &  5.8  & 0.5  &   3.39  & 0.21 &\\
          $f_1+f_0$ &  5&0606936     &  6.3  & 0.5  &   2.82  & 0.16 &\\
$f_1+f_0+\fbj-\fbd$ &  5&0623619     &  2.5  & 0.5  &   3.89  & 0.37 &\\
     $f_1+f_0+\fbd$ &  5&0637907     &  5.6  & 0.5  &   0.90  & 0.18 &\\
     $f_1+f_0+\fbj$ &  5&0654590     &  3.2  & 0.5  &   0.88  & 0.28 &\\
       $2f_1-2\fbj$ &  5&7893134     &  3.6  & 0.5  &   5.74  & 0.27 &\\
  $2f_1-2\fbj+\fbd$ &  5&7924105     &  2.6  & 0.5  &   3.52  & 0.33 &\\
        $2f_1-\fbj$ &  5&7940788     &  4.9  & 0.4  &   3.01  & 0.18 &\\
  $2f_1+\fbj-2\fbd$ &  5&7974155     &  3.1  & 0.4  &   0.50  & 0.49 &\\
             $2f_1$ &  5&7988443     &  7.8  & 0.4  &   0.83  & 0.20 &\\
         $2f_1+f_0$ &  7&9601157     &  3.1  & 0.5  &   4.37  & 0.27 &\\
    $2f_1+f_0+\fbd$ &  7&9632128     &  2.2  & 0.5  &   2.27  & 0.31 & 3.8\\
             $3f_1$ &  8&6982664     &  1.9  & 0.4  &   3.29  & 0.37 & 3.5\\
\hline
\end{tabular}
\end{table*}

\begin{table*}
\centering
\caption{Same as Tab.~\ref{tab:freqs00951} for OGLE-BLG-RRLYR-09284.}
\label{tab:freqs09284}
\begin{tabular}{lr@{.}lrrrrr}
freq. id & \multicolumn{2}{c}{$f$\thinspace [d$^{-1}$]}& $A$\thinspace [mmag] & $\sigma$ & $\phi$ [rad] & $\sigma$ & remarks\\ 
\hline
        $\fbt$ &  0&020970(22) &        &     &       &      & bl\\ 
        $\fbj$ &  0&023935(10) &        &     &       &      & bl\\ 
        $\fbd$ &  0&024537(15) &        &     &       &      & bl\\ 
 $f_1-f_0-\fbj$ &  0&7826980    &  1.4  & 0.2 &  4.05 & 0.40 &\\
      $f_1-f_0$ &  0&8066332    & *6.1  & 0.2 &  3.56 & 0.10 &\\
     $f_0-\fbd$ &  2&2648433    &  3.8  & 0.2 &  2.70 & 0.59 &\\
          $f_0$ &  2&2893799(22)&*25.5  & 0.2 &  4.65 & 0.08 &\\
     $f_0+\fbt$ &  2&3103501    & *2.8  & 0.2 &  5.18 & 0.87 &\\
     $f_0+\fbj$ &  2&3133150    & *6.4  & 0.2 &  5.01 & 0.38 &\\
          $f_1$ &  3&0960130(10)& 60.1  & 0.2 &  2.77 & 0.04 &\\
     $f_{\rm u}$ &  3&170648(47) &  1.3  & 0.2 &  2.96 & 1.82 &\\
         $2f_0$ &  4&5787597    &  2.7  & 0.2 &  5.61 & 0.18 &\\
    $2f_0+\fbj$ &  4&6026949    &  1.2  & 0.2 &  5.86 & 0.41 &\\
 $f_1+f_0-\fbd$ &  5&3608564    &  1.4  & 0.2 &  1.96 & 0.60 &\\
      $f_1+f_0$ &  5&3853929    & *9.8  & 0.2 &  3.86 & 0.09 &\\
 $f_1+f_0+\fbt$ &  5&4063632    &  1.3  & 0.2 &  4.30 & 0.88 &\\
 $f_1+f_0+\fbj$ &  5&4093281    &  2.5  & 0.2 &  4.12 & 0.39 &\\
         $2f_1$ &  6&1920261    &  6.2  & 0.2 &  2.73 & 0.08 &\\
     $3f_1-f_0$ &  6&9986593    &  1.2  & 0.2 &  6.13 & 0.21 &\\
     $2f_0+f_1$ &  7&6747727    &  1.7  & 0.2 &  4.60 & 0.20 &\\
     $2f_1+f_0$ &  8&4814059    &  3.8  & 0.2 &  3.61 & 0.12 &\\
$2f_1+f_0+\fbj$ &  8&5053411    &  1.3  & 0.2 &  3.73 & 0.41 &\\
         $3f_1$ &  9&2880391    &  3.7  & 0.2 &  2.79 & 0.12 &\\
\hline
\end{tabular}
\end{table*}

\begin{table*}
\centering
\caption{Same as Tab.~\ref{tab:freqs00951} for OGLE-BLG-RRLYR-10728.}
\label{tab:freqs10728}
\begin{tabular}{lr@{.}lrrrrr}
freq. id & \multicolumn{2}{c}{$f$\thinspace [d$^{-1}$]}& $A$\thinspace [mmag] & $\sigma$ & $\phi$ [rad] & $\sigma$ & remarks\\ 
\hline
              $\fbd$ &  0&003199(18) &       &     &       &      &bl\\ 
              $\fbj$ &  0&0101305(82)&   5.6  & 0.7 &  3.29 & 0.34 &\\  
           $f_1-f_0$ &  0&7383676    &  11.1  & 0.7 &  4.98 & 0.16 &\\
          $2f_0-f_1$ &  1&3377177    &   7.4  & 0.7 &  3.51 & 0.20 &\\
     $f_0-\fbj-\fbd$ &  2&0627558    &   5.7  & 0.7 &  0.51 & 0.71 &\\
          $f_0-\fbd$ &  2&0728863    &   8.9  & 0.7 &  0.84 & 0.70 &\\
               $f_0$ &  2&0760853(11)&*179.9  & 0.7 &  1.73 & 0.04 &\\
          $f_0+\fbj$ &  2&0862158    &  26.7  & 0.7 &  3.53 & 0.31 &\\
          $f_1-\fbj$ &  2&8043225    &   8.6  & 0.7 &  1.08 & 0.36 &\\
               $f_1$ &  2&8144530(38)& *54.2  & 0.7 &  1.02 & 0.15 &\\
          $3f_0-f_1$ &  3&4138031    &   4.2  & 0.7 &  0.86 & 0.26 &\\
         $2f_0-\fbd$ &  4&1489716    &   4.7  & 0.7 &  5.01 & 0.71 &\\
              $2f_0$ &  4&1521707    & *49.9  & 0.7 &  5.61 & 0.09 &\\
         $2f_0+\fbj$ &  4&1623012    &   6.1  & 0.7 &  1.06 & 0.34 &\\
    $2f_0+\fbj+\fbd$ &  4&1655002    &   5.4  & 0.7 &  4.36 & 0.71 &\\
 $f_0+f_1-\fbj-\fbd$ &  4&8772088    &   4.6  & 0.7 &  3.46 & 0.72 &\\
      $f_0+f_1-\fbj$ &  4&8804078    &   3.8  & 0.7 &  5.60 & 0.41 &\\
           $f_0+f_1$ &  4&8905383    &  25.3  & 0.7 &  5.52 & 0.16 &\\
      $f_0+f_1+\fbj$ &  4&9006688    &   3.8  & 0.7 &  0.59 & 0.40 &\\
              $2f_1$ &  5&6289059    &   5.7  & 0.7 &  4.16 & 0.32 &\\
              $3f_0$ &  6&2282560    &  24.7  & 0.7 &  3.39 & 0.14 &\\
         $3f_0+\fbj$ &  6&2383865    &   3.0  & 0.7 &  4.89 & 0.40 &\\
    $3f_0+\fbj+\fbd$ &  6&2415856    &   3.8  & 0.7 &  1.58 & 0.73 &\\
$2f_0+f_1-\fbj-\fbd$ &  6&9532941    &   4.6  & 0.7 &  2.24 & 0.72 &\\
          $2f_0+f_1$ &  6&9666236    &  14.6  & 0.7 &  3.62 & 0.18 &\\
          $f_0+2f_1$ &  7&7049912    &   7.1  & 0.7 &  3.17 & 0.32 &\\
              $4f_0$ &  8&3043414    &  11.2  & 0.7 &  1.36 & 0.19 &\\
          $3f_0+f_1$ &  9&0427089    &   7.2  & 0.7 &  1.38 & 0.23 &\\
         $2f_0+2f_1$ &  9&7810766    &   5.7  & 0.7 &  1.28 & 0.33 &\\
              $5f_0$ & 10&3804267    &   5.2  & 0.7 &  5.44 & 0.25 &\\
          $4f_0+f_1$ & 11&1187943    &   3.9  & 0.7 &  4.98 & 0.29 &\\
\hline
\end{tabular}
\end{table*}

\begin{table*}
\centering
\caption{Same as Tab.~\ref{tab:freqs00951} for OGLE-BLG-RRLYR-11311.}
\label{tab:freqs11311}
\begin{tabular}{lr@{.}lrrrrr}
freq. id & \multicolumn{2}{c}{$f$\thinspace [d$^{-1}$]}& $A$\thinspace [mmag] & $\sigma$ & $\phi$ [rad] & $\sigma$ & remarks\\ 
\hline
         $\fbd$ &  0&011477(21) &        &      &         &      &bl\\ 
         $\fbj$ &  0&020353(10) &        &      &         &      &bl\\ 
      $f_1-f_0$ &  0&7605726    &   5.9  & 1.0  &   2.58  & 0.35 &\\
          $f_0$ &  2&1348138(66)& *50.3  & 1.0  &   3.73  & 0.25 &\\
   $f_0+\fbd/2$ &  2&1405521    &   6.7  & 1.0  &   3.05  & 0.40 &\\
     $f_0+\fbd$ &  2&1462904    &  13.8  & 1.1  &   0.42  & 0.73 &\\
     $f_1-\fbj$ &  2&8750334    & *32.4  & 1.0  &   2.43  & 0.35 &\\
          $f_1$ &  2&8953863(37)& *91.5  & 1.0  &   5.70  & 0.14 &\\
    $2f_0+\fbd$ &  4&2811041    &  10.5  & 1.0  &   5.92  & 0.74 &\\
 $f_1+f_0-\fbj$ &  5&0098471    &  17.7  & 1.0  &   1.72  & 0.41 &\\
      $f_1+f_0$ &  5&0302001    &   5.1  & 1.0  &   5.10  & 0.35 &\\
    $2f_1-\fbj$ &  5&7704197    &   5.2  & 1.0  &   4.65  & 0.43 &\\
         $2f_1$ &  5&7907726    &  10.4  & 1.0  &   1.68  & 0.30 &\\
$2f_1+f_0-\fbj$ &  7&9052334    &   8.4  & 1.0  &   3.67  & 0.44 &\\
\hline
\end{tabular}
\end{table*}

\begin{table*}
\centering
\caption{Same as Tab.~\ref{tab:freqs00951} for OGLE-BLG-RRLYR-13442.}
\label{tab:freqs13442}
\begin{tabular}{lr@{.}lrrrrr}
freq. id & \multicolumn{2}{c}{$f$\thinspace [d$^{-1}$]}& $A$\thinspace [mmag] & $\sigma$ & $\phi$ [rad] & $\sigma$ & remarks\\ 
\hline
       $\fbj$ & 0&005583(29) &       &     &         &      &bl\\
    $f_1-f_0$ & 0&7674837    &   9.7 & 1.6 &    2.19 & 0.50 &\\
        $f_0$ & 2&0572265(45)&*115.3 & 1.6 &    2.29 & 0.17 &\\
   $f_0+\fbj$ & 2&0628097    & *14.0 & 1.6 &    0.71 & 1.13 &\\
   $f_{\rm u1}$ & 2&085225(19) &  28.8 & 1.6 &    4.12 & 0.75 &\\
   $f_{\rm u2}$ & 2&089231(56) &   9.9 & 1.6 &    1.32 & 2.15 &\\
        $f_1$ & 2&824710(11) & *44.3 & 1.6 &    4.33 & 0.43 &\\
       $2f_0$ & 4&1144530    &  14.2 & 1.6 &    1.29 & 0.36 &\\
  $2f_0+\fbj$ & 4&1200362    &  10.8 & 1.6 &    4.70 & 1.14 &\\
    $f_0+f_1$ & 4&8819367    & *11.8 & 1.6 &    2.70 & 0.48 &\\
       $2f_1$ & 5&6494204    &   6.7 & 1.6 &    5.14 & 0.89 &\\
\hline
\end{tabular}
\end{table*}

\begin{table*}
\centering
\caption{Same as Tab.~\ref{tab:freqs00951} for OGLE-BLG-RRLYR-14915.}
\label{tab:freqs14915}
\begin{tabular}{lr@{.}lrrrrr}
freq. id & \multicolumn{2}{c}{$f$\thinspace [d$^{-1}$]}& $A$\thinspace [mmag] & $\sigma$ & $\phi$ [rad] & $\sigma$ & remarks\\ 
\hline
            $\fbd$ &  0&007135(32) &       &       &        &      &bl\\
            $\fbj$ &  0&0158997(92)&  3.7  & 0.8   &  3.83  & 0.43 & 3.7\\
$f_1-f_0-\fbj-\fbd$ &  0&8240548    &  5.7  & 0.9   &  2.47  & 1.25 &\\
     $f_1-f_0-\fbj$ &  0&8311896    &  8.4  & 0.9   &  4.83  & 0.32 &\\
          $f_1-f_0$ &  0&8470893    &  8.5  & 0.9   &  2.16  & 0.24 &\\
              $f_0$ &  2&3074665(29)&*99.7  & 0.9   &  4.29  & 0.11 &\\
         $f_1-\fbj$ &  3&1386561    &*41.2  & 0.9   &  3.73  & 0.29 &\\
       $f_1-\fbj/2$ &  3&1466052    &  4.8  & 0.9   &  5.96  & 0.23 &\\
              $f_1$ &  3&1545558(41)&*64.1  & 0.9   &  5.86  & 0.16 &\\
             $2f_0$ &  4&6149329    &  9.2  & 0.8   &  5.51  & 0.24 &\\
     $f_0+f_1-\fbj$ &  5&4461225    &  5.4  & 0.9   &  3.89  & 0.35 &\\
          $f_0+f_1$ &  5&4620222    & 27.4  & 0.9   &  5.90  & 0.18 &\\
     $f_0+f_1+\fbd$ &  5&4691571    &  5.7  & 0.8   &  5.49  & 1.23 &\\
        $2f_1-\fbj$ &  6&2932118    &  8.6  & 0.9   &  5.68  & 0.32 &\\
             $2f_1$ &  6&3091115    &  4.9  & 0.8   &  2.22  & 0.36 &\\
   $2f_1+\fbj-\fbd$ &  6&3178764    &  4.5  & 0.8   &  4.32  & 1.41 &\\
         $2f_0+f_1$ &  7&7694887    &  6.2  & 0.8   &  0.46  & 0.29 &\\
    $2f_0+f_1+\fbd$ &  7&7766236    &  3.4  & 0.8   &  6.09  & 1.24 &\\
    $f_0+2f_1-\fbj$ &  8&6006783    &  3.9  & 0.9   &  5.28  & 0.39 &\\
         $f_0+2f_1$ &  8&6165780    &  7.8  & 0.8   &  1.77  & 0.34 &\\
        $2f_0+2f_1$ & 10&9240445    &  5.0  & 0.8   &  2.06  & 0.40 &\\
   $2f_0+2f_1+\fbd$ & 10&9311793    &  4.3  & 0.9   &  1.86  & 1.26 &\\
\hline
\end{tabular}
\end{table*}

\begin{table*}
\centering
\caption{Same as Tab.~\ref{tab:freqs00951} for OGLE-BLG-RRLYR-22356.}
\label{tab:freqs22356}
\begin{tabular}{lr@{.}lrrrrr}
freq. id & \multicolumn{2}{c}{$f$\thinspace [d$^{-1}$]}& $A$\thinspace [mmag] & $\sigma$ & $\phi$ [rad] & $\sigma$ & remarks\\ 
\hline
        $\fbj$  & 0&0039145(73) &   5.0  & 1.0   & 0.28  & 0.35 &\\  
        $\fbd$  & 0&019384(23)  &        &       &       &      &bl\\
        $\fbt$  & 0&030803(34)  &        &       &       &      &bl\\
$f_1-f_0+\fbj$  & 0&8343478     &   6.4  & 1.0   & 3.45  & 0.58 &\\
$f_1-f_0+\fbd$  & 0&8498175     &   6.3  & 1.0   & 0.14  & 0.78 &\\
    $f_0-\fbt$  & 2&2284080     &   7.4  & 1.0   & 5.99  & 1.32 &\\
    $f_{\rm u}$  & 2&243391(26)  & *13.1  & 1.0   & 1.39  & 0.98 &\\
    $f_0-\fbj$  & 2&2552965     &  40.1  & 1.0   & 5.19  & 0.27 &\\
         $f_0$  & 2&2592110(22) &*147.5  & 1.0   & 4.26  & 0.09 &\\
         $f_1$  & 3&089644(12)  &  21.9  & 1.0   & 3.38  & 0.48 &\\
    $f_1+\fbd$  & 3&1090285     & *15.0  & 1.0   & 5.28  & 0.76 &\\
   $2f_0-\fbt$  & 4&4876190     &   6.4  & 1.0   & 6.19  & 1.32 &\\
  $2f_0-2\fbj$  & 4&5105930     &  *9.9  & 1.0   & 5.45  & 0.55 &\\
   $2f_0-\fbj$  & 4&5145075     &   8.3  & 1.0   & 0.03  & 0.31 &\\
        $2f_0$  & 4&5184220     & *18.6  & 1.0   & 4.69  & 0.18 &\\
$f_0+f_1-\fbj$  & 5&3449408     &   9.8  & 1.0   & 3.96  & 0.54 &\\
     $f_0+f_1$  & 5&3488553     &   5.3  & 1.0   & 3.99  & 0.51 &\\
$f_0+f_1+\fbd$  & 5&3682395     &  *4.7  & 1.0   & 5.74  & 0.79 &\\
    $f_0+2f_1$  & 8&4384996     &   4.4  & 1.0   & 3.06  & 0.98 &\\
\hline
\end{tabular}
\end{table*}

\begin{table*}
\centering
\caption{Same as Tab.~\ref{tab:freqs00951} for OGLE-BLG-RRLYR-30986.}
\label{tab:freqs30986}
\begin{tabular}{lr@{.}lrrrrr}
freq. id & \multicolumn{2}{c}{$f$\thinspace [d$^{-1}$]}& $A$\thinspace [mmag] & $\sigma$ & $\phi$ [rad] & $\sigma$ & remarks\\ 
\hline
        $\fbj$ &  0&0089803(84) &       &      &         &      &bl\\
$f_1-f_0-\fbj$ &  0&7224397     &  9.0  & 1.3  &   0.04  & 0.46 & \\
     $f_1-f_0$ &  0&7314200     &  7.7  & 1.3  &   2.05  & 0.32 & \\
    $f_0-\fbj$ &  2&0141860     & 51.0  & 1.3  &   6.20  & 0.30 & \\
         $f_0$ &  2&0231663(43) & 63.8  & 1.3  &   2.21  & 0.17 & \\
   $f_1-2\fbj$ &  2&7366258     & 12.3  & 1.3  &   6.13  & 0.65 & \\
    $f_1-\fbj$ &  2&7456060     & 14.4  & 1.3  &   3.54  & 0.37 & \\
         $f_1$ &  2&7545863(60) & 73.6  & 1.3  &   3.43  & 0.23 & \\
        $2f_0$ &  4&0463326     &*37.0  & 1.3  &   5.96  & 0.33 & \\
$f_1+f_0-\fbj$ &  4&7687723     &  7.0  & 1.3  &   5.72  & 0.38 & \\
     $f_1+f_0$ &  4&7777525     &  7.2  & 1.3  &   1.23  & 0.34 & \\
        $2f_1$ &  5&5091725     &  8.7  & 1.3  &   3.29  & 0.48 & \\
        $3f_0$ &  6&0694988     &  8.0  & 1.3  &   4.81  & 0.51 & \\
    $f_1+2f_0$ &  6&8009188     &  9.9  & 1.3  &   5.74  & 0.43 & \\
\hline
\end{tabular}
\end{table*}

\begin{table*}
\centering
\caption{Same as Tab.~\ref{tab:freqs00951} for OGLE-BLG-RRLYR-32462.}
\label{tab:freqs32462}
\begin{tabular}{lr@{.}lrrrrr}
freq. id & \multicolumn{2}{c}{$f$\thinspace [d$^{-1}$]}& $A$\thinspace [mmag] & $\sigma$ & $\phi$ [rad] & $\sigma$ & remarks\\ 
\hline
         $\fbj$ & 0&0052617(81) &       &       &       &      &bl\\  
         $\fbd$ & 0&007959(11)  &       &       &       &      &bl\\  
 $f_1-f_0-\fbd$ & 0&8004467     &  4.6  & 0.8   & 1.43  & 0.44 & \\
      $f_1-f_0$ & 0&8084058     &  7.6  & 0.9   & 2.61  & 0.22 & \\
 $f_1-f_0+\fbj$ & 0&8136674     &  4.8  & 0.9   & 6.12  & 0.38 & \\
          $f_0$ & 2&2947122(38) &*67.6  & 0.8   & 3.37  & 0.14 & \\
     $f_0+\fbd$ & 2&3026713     & 23.5  & 0.8   & 4.48  & 0.38 & \\
     $f_1-\fbd$ & 3&0951589     &  8.8  & 0.8   & 0.53  & 0.43 & \\
          $f_1$ & 3&1031180(31) &*82.7  & 0.8   & 5.47  & 0.12 & \\
     $f_1+\fbj$ & 3&1083797     &*31.4  & 0.9   & 2.76  & 0.29 & \\
         $2f_0$ & 4&5894245     & *5.2  & 0.8   & 3.03  & 0.33 & \\
      $f_1+f_0$ & 5&3978302     &*11.5  & 0.8   & 5.49  & 0.20 & \\
 $f_1+f_0+\fbj$ & 5&4030919     &*14.9  & 1.1   & 1.70  & 0.31 & \\
 $f_1+f_0+\fbd$ & 5&4057893     & 10.9  & 1.1   & 5.86  & 0.41 & \\
         $2f_1$ & 6&2062360     & 11.7  & 0.8   & 1.25  & 0.24 & \\
    $2f_1+\fbj$ & 6&2114976     &  5.0  & 0.8   & 4.50  & 0.35 & \\
$f_1+2f_0+\fbj$ & 7&6978041     &  5.7  & 0.8   & 1.82  & 0.40 & \\
     $2f_1+f_0$ & 8&5009482     &  6.0  & 0.8   & 1.70  & 0.30 & \\
$2f_1+f_0+\fbj$ & 8&5062099     &  8.7  & 0.8   & 3.78  & 0.34 & \\
         $3f_1$ & 9&3093540     &  4.9  & 0.8   & 2.72  & 0.39 & \\
\hline
\end{tabular}
\end{table*}

\label{lastpage}


\begin{thebibliography}{99}

\bibitem[\protect\citeauthoryear{Alcock et al.}{2003}]{alcock_rrab} Alcock C., et al., 2003, ApJ, 598, 597
\bibitem[\protect\citeauthoryear{Benk\H{o}, Szab\'o \& Papar\'o}{2011}]{benko11} Benk\H{o} J.M., Szab\'o R.., Papar\'o M., 2011, MNRAS, 417, 974
\bibitem[\protect\citeauthoryear{Benk\H{o} et al.}{2010}]{benko10} Benk\H{o} J.M., Kolenberg K., Szab\'o R. et al., 2010, MNRAS, 409, 1585
\bibitem[\protect\citeauthoryear{Benk\H{o} et al.}{2014}]{benko14} Benk\H{o} J.M., Plachy E., Szab\'o R. et al., 2014, ApJ Suppl. Ser., 213, 131
\bibitem[\protect\citeauthoryear{Buchler \& Koll\'ath}{2011}]{bk11}  Buchler J.R., Koll\'ath Z., 2011, ApJ Lett., 731, 24
\bibitem[\protect\citeauthoryear{Blazhko}{1907}]{blazhko} Blazhko S.N., 1907, Astron. Nachr., 175, 325
\bibitem[\protect\citeauthoryear{Chadid}{2012}]{chadid12} Chadid M., 2012, A\&A, 540, A68
\bibitem[\protect\citeauthoryear{Chadid et al.}{2010}]{chadid10} Chadid M. et al., 2010, A\&A, 510, A39
\bibitem[\protect\citeauthoryear{Dziembowski}{2012}]{wd12} Dziembowski W., 2012, Acta Astron., 62, 323
\bibitem[\protect\citeauthoryear{Gruberbauer et al.}{2007}]{aqleo} Gruberbauer M., Kolenberg K., Rowe J. et al., 2007, MNRAS, 379, 1498
\bibitem[\protect\citeauthoryear{Guggenberger et al.}{2012}]{gugg} Guggenberger E., et al., 2012, MNRAS, 424, 649
\bibitem[\protect\citeauthoryear{Jurcsik et al.}{2008}]{jurcsik_mwlyr} Jurcsik J., et al., 2008, MNRAS, 391, 164
\bibitem[\protect\citeauthoryear{Jurcsik et al.}{2009}]{jurcsik_kbs} Jurcsik J., S\'odor A., Szeidl B. et al., 2009, MNRAS, 400, 1006
\bibitem[\protect\citeauthoryear{Jurcsik et al.}{2014}]{jurcsik_bl} Jurcsik J., Smitola, P., Hajdu G., Nuspl J., 2014, ApJ Lett., 797, L3
\bibitem[\protect\citeauthoryear{Kov\'acs, Buchler \& Davis}{1987}]{kbd87}  Kov\'acs G., Buchler J.R., Davis C.G., 1987, ApJ, 319, 247
\bibitem[\protect\citeauthoryear{Le Borgne et al.}{2014}]{leborgne}  Le Borgne J.F., et al., 2014, MNRAS, 441, 1435
\bibitem[\protect\citeauthoryear{Mizerski}{2003}]{mizerski} Mizerski T., 2003, Acta Astron., 53, 307
\bibitem[\protect\citeauthoryear{Moln\'ar et al.}{2012}]{molnar12} Moln\'ar L., Koll\'ath Z., Szab\'o R., 2012, ApJ Lett., 757, L13
\bibitem[\protect\citeauthoryear{Moskalik \& Poretti}{2003}]{mp03} Moskalik P., Poretti E., 2003, A\&A, 398, 213
\bibitem[\protect\citeauthoryear{Moskalik \& Ko\l{}aczkowski}{2009}]{mk09} Moskalik P., Ko\l{}aczkowski Z., 2009, MNRAS, 394, 1649
\bibitem[\protect\citeauthoryear{Moskalik et al.}{2015}]{pamsm14} Moskalik P.,  Smolec R. \& Kolenberg K., et al., 2015, MNRAS, in press, arXiv:1412.2272
\bibitem[\protect\citeauthoryear{Nagy \& Kov\'acs}{2006}]{nagy} Nagy A., Kov\'acs G., 2006, A\&A, 454, 257
\bibitem[\protect\citeauthoryear{Netzel, Smolec \& Moskalik}{2014}]{netzel} Netzel H., Smolec R. \& Moskalik P., 2014, MNRAS, in press, arXiv:1411.3155
\bibitem[\protect\citeauthoryear{Olech \& Moskalik}{2009}]{om09} Olech A., Moskalik P., 2009, A\&A, 494, L17
\bibitem[\protect\citeauthoryear{Petersen}{1978}]{petersen} Petersen J.O. 1978, A\&A, 62, 205
\bibitem[\protect\citeauthoryear{Pietrzy\'nski et al.}{2012}]{gp12} Pietrzy\'nski G. et al., 2012, Nature, 484, 75
\bibitem[\protect\citeauthoryear{Popielski, Dziembowski \& Cassisi}{2000}]{popielski} Popielski B.L., Dziembowski W., Cassisi S., 2000, Acta Astron., 50, 491
\bibitem[\protect\citeauthoryear{Skarka}{2013}]{skarka} Skarka M., 2013, A\&A, 549, 101
\bibitem[\protect\citeauthoryear{Smolec}{2014}]{smolec14} Smolec R., 2014, IAUS, 301, 265
\bibitem[\protect\citeauthoryear{Smolec et al.}{2013}]{smolec_bep} Smolec R., Pietrzy\'nski G., Graczyk D., et al., 2013, MNRAS, 428, 3034
\bibitem[\protect\citeauthoryear{Smolec et al.}{2014}]{smolec_tri} Smolec R., Soszy\'nski I., Udalski A., et al., 2014, MNRAS, submitted, arXiv:1411.2908
\bibitem[\protect\citeauthoryear{Soszy\'nski et al.}{2009}]{ogle_rr_lmc} Soszy\'nski I., Udalski A., Szyma\'nski M.K. et al., 2009, Acta Astron., 59, 1
\bibitem[\protect\citeauthoryear{Soszy\'nski et al.}{2010a}]{ogle_cep_smc} Soszy\'nski I., Poleski R., Udalski A., et al., 2010, Acta Astron., 60, 17
\bibitem[\protect\citeauthoryear{Soszy\'nski et al.}{2010b}]{ogle_rr_smc} Soszy\'nski I., Udalski A., Szyma\'nski M.K. et al., 2010, Acta Astron., 60, 165
\bibitem[\protect\citeauthoryear{Soszy\'nski et al.}{2011}]{ogle_rrl_blg} Soszy\'nski I., Dziembowski W., Udalski A., et al., 2011, Acta Astron., 61, 1
\bibitem[\protect\citeauthoryear{Soszy\'nski et al.}{2014a}]{ogle_switch} Soszy\'nski I., Dziembowski W., Udalski A., et al., 2014, Acta Astron., 64, 1
\bibitem[\protect\citeauthoryear{Soszy\'nski et al.}{2014b}]{ogleiv_rrl_blg} Soszy\'nski I., Udalski A., Szyma\'nski M.K., et al., 2014, Acta Astron., 64, 177
\bibitem[\protect\citeauthoryear{Szab\'o}{2014}]{szabo14} Szab\'o R., 2014, IAUS, 301, 241
\bibitem[\protect\citeauthoryear{Szab\'o et al.}{2010}]{szabo10} Szab\'o R., Koll\'ath Z., Moln\'ar L. et al., 2010, MNRAS, 409, 1244
\bibitem[\protect\citeauthoryear{Szab\'o et al.}{2014}]{szabo_corot} Szab\'o R., Benk\H{o} J.M., Papar\'o M., 2014, A\&A, 570, A100
\bibitem[\protect\citeauthoryear{Udalski et al.}{2008}]{ogleIII} Udalski A., Szyma\'nski M.K., Soszy\'nski I., Poleski R., 2008, Acta Astron., 58, 69
\end{thebibliography}
\end{document}